\documentclass[tightenlines,superscriptaddress,showpacs,showkeys,preprint,
floatfix]{revtex4}
\usepackage{graphicx} 
\usepackage{overpic} 
 
\begin{document}

\title{\bf Strange Atoms, Strange Nuclei and Kaon Condensation\footnote{
Lectures delivered by Avraham Gal at the International School of Physics 
Enrico Fermi, Varenna, June 2007.}}

\date{\today} 

\author{E.~Friedman}
\email{elifried@vms.huji.ac.il}
\affiliation{Racah Institute of Physics, The Hebrew University,
Jerusalem 91904, Israel\vspace*{1ex}}

\author{A.~Gal}
\email{avragal@vms.huji.ac.il}
\affiliation{Racah Institute of Physics, The Hebrew University,
Jerusalem 91904, Israel\vspace*{1ex}}

\begin{abstract}

Analyses of strong--interaction data consisting of level shifts, 
widths and yields in strange atoms of $K^-$ mesons and $\Sigma^-$ 
hyperons are reviewed. Recent results obtained by fitting to 
comprehensive sets of data across the periodic table in terms of 
density dependent optical potentials are discussed. 
The introduction of density dependence generally improves 
significantly the fit to the data, leading to novel results on the 
in-medium hadron-nucleon $t$ matrix $t(\rho)$ over a wide range of 
densities up to central nuclear densities. A strongly attractive 
$K^-$--nuclear potential of order 150--200 MeV in nuclear matter is 
suggested by fits to $K^-$--atom data, with interesting possible 
repercussions on $\bar K$ condensation and on the evolution of 
strangeness in high-density stars. 
The case for relatively narrow deeply bound $K^-$ {\it atomic} 
states is made, essentially independent of the $K^-$ potential depth. 
In view of the recently reported inconclusive experimental signals of 
$\bar K$ deeply bound states, dynamical models for calculating 
binding energies and widths of $\bar K$-{\it nuclear} states are discussed. 
Lower bounds on the width, $\Gamma_{\bar K} \gtrsim 50$~MeV, are established.  
For $\Sigma^-$ atoms, the fitted potential becomes repulsive inside 
the nucleus, in agreement with recently reported $(\pi^-,K^+)$ spectra 
from KEK, implying that $\Sigma$ hyperons generally do not bind in nuclei. 
This repulsion significantly affects calculated compositions and masses 
of neutron stars.

\end{abstract}



\maketitle 

\tableofcontents 

\hfil\eject 

\section{Introduction} 
\label{sec:intro}

An exotic atom is formed when a negatively charged particle stops 
in a target and is captured by a target atom into an outer atomic 
orbit. It will then emit Auger electrons and characteristic X-rays 
whilst cascading down its own sequence of atomic levels until, 
at some state of low principal quantum number $n$,
the particle is absorbed due to its interaction with the nucleus.
The lifetimes of the particles considered here, namely $K^-$ and 
$\Sigma^-$, are much longer than typical slowing down times and 
atomic time scales. Therefore, following the stopping of the hadron 
in matter, well defined states of an exotic atom are established and 
the effects of the hadron-nucleus strong interaction can be studied. 
The overlap of the atomic orbitals with the nucleus covers a wide 
range of nuclear densities thus creating a unique source of 
information on the density dependence of the hadronic interaction. 

In the study of strong interaction effects in exotic atoms, 
the observables of interest are the shifts ($\epsilon$) and widths 
($\Gamma$) of the atomic levels caused by the strong interaction 
with the nucleus. These levels are shifted and broadened relative 
to the electromagnetic case but the shifts and widths can usually 
only be measured directly for one, or possibly two levels in any 
particular hadronic atom. 
The broadening due to the nuclear absorption usually terminates the 
atomic cascade at low $n$ thus limiting the experimentally observed 
X-ray spectrum. In some cases the width of the next higher $n+1$ 
`upper' level can be obtained indirectly from measurements of the 
relative yields of X-rays when they  depart from their purely 
electromagnetic values. As the atomic number and size of the nucleus 
increase, so the absorption occurs from higher $n$-values as shown 
for $K^-$ atoms and for $\Sigma^-$ atoms in the corresponding Sections. 
Shifts and widths caused by the interaction with the nucleus may be 
calculated by adding an optical potential to the Coulomb interaction. 
The study of the strong interaction in exotic atoms thus becomes the 
study of this additional potential, as reviewed in great detail by Batty, 
Friedman and Gal \cite{BFG97} and very recently by Friedman and Gal 
\cite{FGa07}. On the experimental side, studies of strong interaction 
effects in exotic atoms have been transformed over the years with the 
introduction of increasingly more advanced X-ray detectors and with 
increasing the efficiency of stopping the hadrons, such as with 
a cyclotron trap \cite{Got04}. 

The present Lectures focus particularly on the physics of the strong 
interaction which can be deduced by studying strange atoms, a term which 
is used for exotic atoms formed initially by stopping $K^-$ mesons in 
matter. The importance of the Strange Atoms subject stems from the progress 
made in recent years in quantifying medium modification effects on the 
hadron-nucleus interaction which has enabled one to achieve improved fits to 
existing data within the framework of commonly accepted models \cite{FGa07}. 
These modified interactions obey a low density limit which has not always 
been enforced in earlier analyses, since it is not always relevant to the 
higher density regime explored, where new features of the hadron nucleus 
interaction may become significant to other fields such as astrophysics.

In the next section we will outline the methodology of exotic-atom studies, 
including common tools such as wave equations and optical potentials. 
Of prime importance is the dependence of the optical potential on the model 
of nuclear density used, with emphasis placed on the overlap region between 
the atomic state studied and the nucleus. Radial sensitivity will be defined, 
to serve as a guide. Finally the extreme non perturbative nature of exotic 
atoms will be discussed. Readers who are not concerned about these tools of 
analyzing exotic atoms may skip the next section and go immediately to the 
sections dealing with $K^-$ atoms and nuclei and with $\Sigma$ hyperons. 
 
\section{Exotic atom methodology} 
\label{sec:meth} 
\subsection{Wave equations and optical potentials} 
\label{sec:pot} 

The interaction of hadrons at threshold with the nucleus is customarily 
described by a Klein-Gordon (KG) equation which for exotic-atom applications 
is of the following form:

\begin{equation}\label{eq:KG} 
\left[ \nabla^2  - 2{\mu}(B+V_{{\rm opt}} + V_c) + (V_c+B)^2\right] \psi = 0~~
~~(\hbar = c = 1) 
\end{equation} 
where $\mu$ is the hadron-nucleus reduced mass, $B$ is the complex binding 
energy and $V_c$ is the finite-size Coulomb interaction of the hadron with 
the nucleus, including vacuum-polarization terms, added according to the 
minimal substitution principle $E \to E - V_c$. A term $2V_cV_{\rm opt}$ 
and a term $2BV_{\rm opt}$ were neglected in Eq.~(\ref{eq:KG}) with respect 
to $2{\mu}V_{\rm opt}$; the term $2BV_{\rm opt}$ has to be reinstated in 
studies of deeply-bound states. The optical potential 
$V_{\rm opt}$ is of the $t\rho(r)$ generic class which for underlying 
$s$-wave hadron-nucleon interactions assumes the form: 

\begin{equation}\label{eq:Vopt} 
2\mu V_{\rm opt}(r) = - 4\pi(1+\frac{A-1}{A}\frac{\mu}{M})
\{b_0[\rho_n(r)+\rho_p(r)] + \tau_z b_1[\rho_n(r)-\rho_p(r)] \} \;. 
\end{equation} 
Here, $\rho_n$ and $\rho_p$ are the neutron and proton density distributions 
normalized to the number of neutrons $N$ and number of protons $Z$, 
respectively, $M$ is the mass of the nucleon and $\tau_z = +1$ for the 
negatively charged hadrons considered in the present Lectures. 
In the impulse approximation, $b_0$ and $b_1$ are minus the hadron-nucleon 
isoscalar and isovector scattering lengths, respectively, which are complex 
for the absorptive strong interactions of $K^-$ mesons and $\Sigma^-$ 
hyperons. Generally these `one-nucleon' parameters are functions of the 
density $\rho$, but often the density dependence may be approximated by 
fitting effective values for $b_0$ and $b_1$ to low-energy data. The 
extension of the threshold KG equation (\ref{eq:KG}) and the optical 
potential (\ref{eq:Vopt}) for scattering problems is straightforward 
\cite{BFG97,FGa07}. 

The use of the KG equation rather than the Dirac equation for Fermions, 
such as $\Sigma$ hyperons, is numerically justified when fine-structure 
effects are negligible or are treated in an average way, as for the X-ray 
transitions considered here.
The leading $j$ dependence ($j = l \pm \frac{1}{2}$) of the energy for 
solutions of the Dirac equation for a point-charge $1/r$ potential 
goes as $(j + \frac{1}{2})^{-1}$, and on averaging it over the projections 
of $j$ gives rise to $(l + \frac{1}{2})^{-1}$ which is precisely the 
leading $l$ dependence of the energy for solutions of the KG equation. 
The higher-order contributions to the spin-orbit splitting are suppressed 
by O$(Z \alpha /n)^2$ which is of order 1$\%$ for the high-$n$ X-ray 
transitions encountered for $\Sigma$ hyperons. This is considerably smaller 
than the experimental errors placed on the measured X-ray transition energies 
and widths.

\subsection{Nuclear densities} 
\label{sec:nucldens} 

The nuclear densities are an essential ingredient of the optical potential. 
The density distribution of the protons is usually considered known as
it is obtained from the nuclear charge distribution by
unfolding the finite size of the charge of the proton. The neutron 
distributions are, however, generally not known to sufficient accuracy. 
For many nuclei there is no direct experimental information whatsoever 
on neutron densities and one must then rely on models which sometimes give 
conflicting results for the root-mean-square (rms) radii. Given this 
unsettled state of affairs, a semi-phenomenological approach was adopted 
that covers a broad range of possible neutron density distributions.

Experience with pionic atoms showed \cite{FGa07} that the feature of 
neutron density distributions which is most relevant in determining 
strong interaction effects in pionic atoms is the radial extent, as 
represented e.g. by $r_n$, the neutron density rms radius. 
A linear dependence of $r_n-r_p$ on $(N-Z)/A$ has been successfully 
employed in $\bar p$ studies \cite{TJL01,JTL04,FGM05}, namely 
\begin{equation} \label{eq:RMF} 
r_n-r_p = \gamma \frac{N-Z}{A} + \delta \; ,
\end{equation}
with $\gamma$ close to 1.0~fm and $\delta$ close to zero. 
Expression (\ref{eq:RMF}) has been 
adopted in analyzing strange atoms and, for lack of better global information
about neutron densities, the value of $\gamma$ was varied over a reasonable
range in fitting to the data. This procedure is based on the expectation 
that for a large data set over the whole of the periodic table some local 
variations will cancel out and that an average behavior may be established. 
Phenomenological studies of in-medium nuclear interactions are based on 
such averages. 

In order to allow for possible differences in the shape of the neutron
distribution, a two-parameter Fermi (2pF) distribution was used both for 
the known proton (unfolded from the charge distribution) and for the 
unknown neutron density distributions 
\begin{equation} 
\label{eq:2pF} 
\rho_{n,p}(r)  = \frac{\rho_{0n,0p}}{1+{\rm exp}((r-R_{n,p})/a_{n,p})} \; ,
\end{equation} 
in the `skin' form of Ref.~\cite{TJL01}. In this form, the same diffuseness 
parameter for protons and neutrons, $a_n=a_p$, is assumed and the $R_n$ 
parameter is determined from the rms radius $r_n$ deduced from 
Eq.~(\ref{eq:RMF}) where $r_p$ is considered to be known. It was checked 
that for $K^-$ atoms and for $\Sigma^-$ atoms the assumption of `skin' 
form was as good, or better, than assuming other (notably `halo') forms. 

Another sensitivity that may be checked in global fits is to the radial 
extension of the hadron-nucleon interaction when folded together with 
the nuclear density. The resultant `finite range' density is defined as 
\begin{equation}
\label{eq:fold} 
\rho ^{\rm F}(r)~~=~~\int d{\bf r}' \rho({\bf r}') \frac{1}{\pi ^{3/2} \beta^3}
e^{-({\bf r}-{\bf r'})^2/\beta^2}~~,
\end{equation}
assuming a Gaussian interaction. It was found that $K^-$ and $\Sigma^-$ atoms 
do not display sensitivity to finite-range effects.

\subsection{Radial sensitivity in exotic atoms} 
\label{sec:fd} 

The radial sensitivity of exotic atom data was addressed before
\cite{BFG97} with the help of a `notch test', introducing a local
perturbation into the potential and studying the changes in the
fit to the data as function of position of the perturbation. The
results gave at least a semi-quantitative information on what are
the radial regions which are being probed by the various types of
exotic atoms. However, the radial extent of the perturbation
was somewhat arbitrary and only very recently that approach was
extended \cite{BFr07} into a mathematically well-defined limit.

In order to study the radial sensitivity of {\it global}
fits to exotic atom data, it is necessary to define the radial position 
parameter globally using as reference, e.g. the known charge distribution 
for each nuclear species in the data base. The radial position $r$
is then defined as $r=R_c+\eta a_c$, where $R_c$ and $a_c$ are the radius 
and diffuseness parameters, respectively, of a 2pF charge 
distribution \cite{FBH95}. In that way $\eta$ becomes the relevant radial 
parameter when handling together data for several nuclear species along 
the periodic table. The value of $\chi ^2$ is regarded now as a functional
of a global optical potential $V(\eta)$, i.e. $\chi ^2=\chi^2[V(\eta)]$,
where the parameter $\eta$ is a {\it continuous} variable.
It leads to \cite{BFr07}
\begin{equation} 
\label{eq:dchi2} 
d\chi^2 = \int d\eta \frac{\delta \chi^2}{\delta V(\eta)} \delta V(\eta) \;,
\end{equation}
where
\begin{equation}
\label{eq:FD}
\frac{\delta \chi^2[V(\eta)]}{\delta V(\eta')}
= \lim_{\sigma \rightarrow 0}\lim_{\epsilon _V \rightarrow 0 }
\frac{\chi^2[V(\eta)+\epsilon _V\delta_{\sigma}(\eta-\eta')]-\chi^2[V(\eta)]}
     {\epsilon _V}\;
\end{equation}
is the functional derivative (FD) of $\chi^2[V]$.
The notation $\delta_{\sigma}(\eta-\eta')$ stands for an approximated
$\delta$-function and $\epsilon _V $ is a change in the potential.
From Eq.~(\ref{eq:dchi2}) it is seen that the FD determines 
the effect of a local
change in the optical potential on $\chi^2$. Conversely it can be said that
the optical potential sensitivity to the experimental data is determined by
the magnitude of the FD.
Calculation of the FD may be carried out by multiplying the
best fit potential by a factor
\begin{equation}
\label{eq:FDfac}
 f=1+\epsilon \delta_{\sigma}(\eta-\eta')
\end{equation}
using a normalized Gaussian with a range parameter $\sigma$ for the
smeared $\delta$-function,
\begin{equation}
\label{eq:Gauss}
\delta_{\sigma}(\eta-\eta')=\frac{1}{\sqrt{2\pi}\sigma}
e^{-(\eta-\eta')^2/2\sigma^2}.
\end{equation}
For finite values of $\epsilon$ and $\sigma$ the FD can 
then be approximated by
\begin{equation}
\label{eq:FDnum}
\frac{\delta \chi^2[V(\eta)]}{\delta V(\eta')}
\approx \frac{1}{V(\eta')}
\frac{\chi^2[V(\eta)(1+\epsilon\delta_{\sigma}(\eta-\eta'))]
      -\chi^2[V(\eta)]}
     {\epsilon}\;.
\end{equation}
The parameter $\epsilon $ is used for a {\it fractional}
change in the potential
and the limit $\epsilon \to 0$ is obtained numerically for several
values of $\sigma $ and then extrapolated to $\sigma =0$.

\subsection{Nonperturbative aspects of exotic atoms} 
\label{sec:zel} 

The optical potentials used to calculate the shifts and widths of atomic 
energy levels are confined to a small region of the atom and thus lead to 
very small energy shifts compared to the corresponding binding energies. 
However, they greatly modify the wavefunction locally and this modification 
makes the strong interaction effects non-perturbative. For example, in kaonic 
atoms all the measured level shifts are repulsive, yet `$t_{\rm eff}\rho$' 
fits to the data invariably lead to attractive potentials. The net repulsive 
shift is due to the imaginary part of the potential being comparable in 
magnitude to the real part. Repulsive shifts may also arise from dominantly 
real attractive optical potentials that are sufficiently strong to bind 
{\it nuclear} states. A nuclear state generated by the optical potential 
gives rise to a node inside the nucleus for the atomic wavefunction by 
orthogonality. The strong modification of atomic wavefunctions due to 
binding nuclear states by $V_{\rm opt}$ may also give rise to irregularities 
in the parameters of $V_{\rm opt}$ obtained from fits to data. This effect was 
first observed by Krell \cite{Kre71} for kaonic atoms. These irregularities 
can be explained by large variations in the atomic wavefunctions such that 
additional nodes may be accommodated within the nucleus. 
A more comprehensive discussion of these nonperturbative aspects is found in 
Ref.~\cite{GFB96}. 

\begin{figure}[t] 
\includegraphics[scale=0.7]{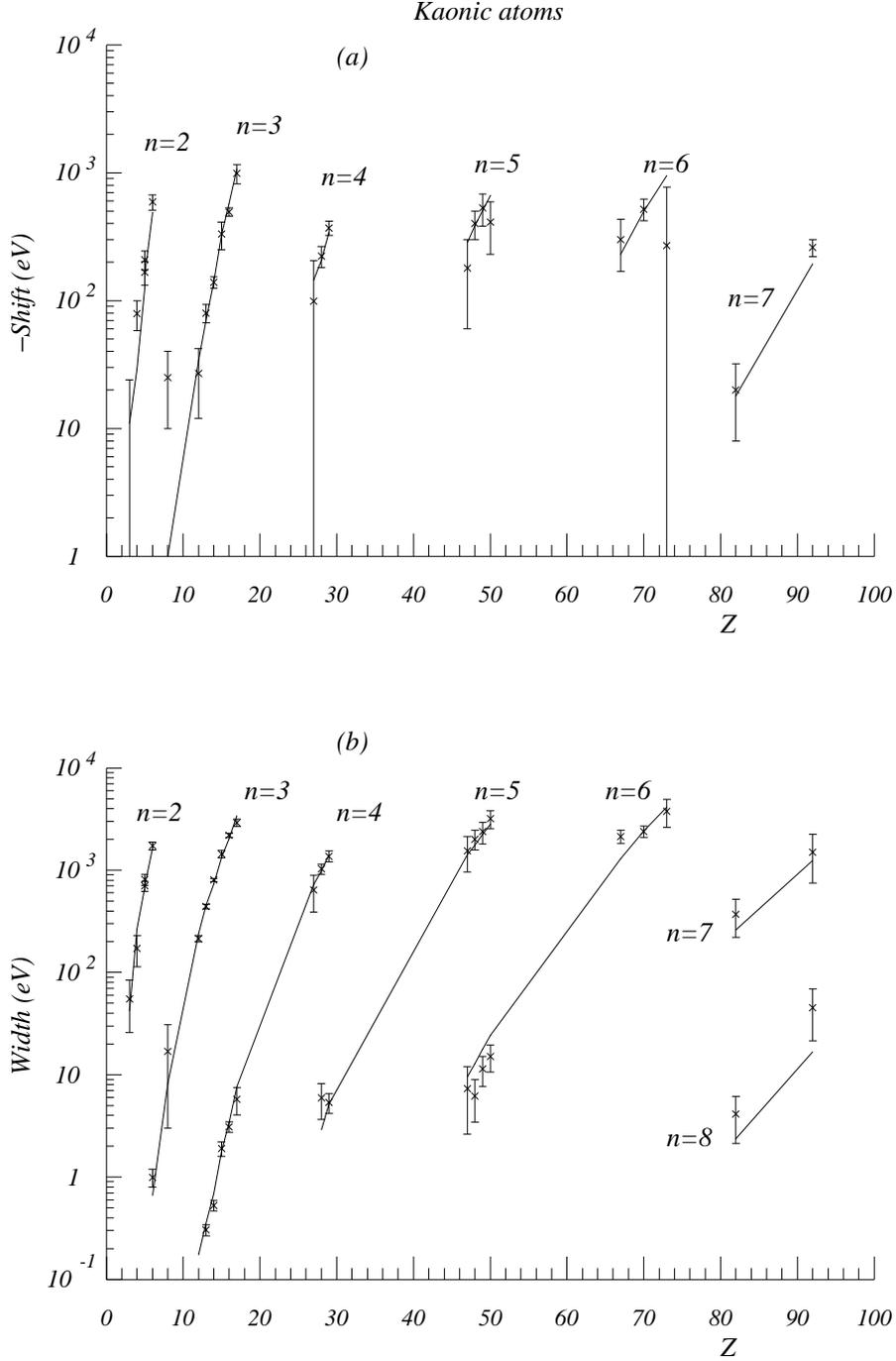} 
\caption{Shift and width values for kaonic atoms. The continuous lines 
join points calculated with a best-fit DD optical potential.} 
\label{fig:katoms} 
\end{figure}

\section{$\bf{K^-}$ Atoms} 
\label{sec:katoms} 

\subsection{Fits to $K^-$ -atom data}
\label{sec:Kat}

The $K^-$-atom data used in global fits~\cite{BFG97} are shown in 
Fig.~\ref{fig:katoms}, spanning a range of atomic states from $2p$ 
in Li to $8j$ in U, with 65 level-shifts, widths and transition yields 
data points. We note that the shifts are `repulsive', largely due 
to the substantial absorptivity of the $K^-$-nuclear interaction. 
It was shown already in the mid 1990s~\cite{BFG97} that 
although a reasonably good fit to the data is obtained 
for a $t\rho $ potential, Eq.~(\ref{eq:Vopt}), with an effective
complex parameter $b_0$ corresponding to attraction, greatly improved fits 
are obtained with a density-dependent potential, where the fixed $b_0$ 
is replaced by 
\begin{equation}
\label{eq:DD} 
b_0+B_0 [\rho(r) /\rho_0 ]^\alpha \;, 
\end{equation} 
with $b_0, B_0$ and $\alpha \geq 0$ determined by fits to the data. 
Fitted potentials of this kind are marked DD. This parameterization 
offers the advantage of fixing $b_0$ at its (repulsive) free-space value 
in order to respect the low-density limit, while relegating the expected 
in-medium attraction to the $B_0$ term which goes with a higher power 
of the density. 

The departure of the optical potential from the fixed-$t$ $t \rho $
approach was recently given a more intrinsically geometrical meaning 
within a model \cite{MFG06} where, loosely speaking, $V_{\rm opt}$ follows 
the shape of a function $F(r)$ inside, and the shape of $[1-F(r)]$ 
outside the nucleus: 
\begin{equation}
\label{eq:DDF}
b_0~\rightarrow ~B_0~F(r)~+~b_0~[1~-~F(r)],~~~~~~~~F(r)~=~\frac{1}{e^x +1}\;, 
\end{equation} 
with $x~=~(r-R_x)/a_x$. Then clearly $F(r)~\rightarrow~1$ for 
$(R_x - r) >> a_x$, which defines the internal region. 
Likewise $[1~-~F(r)]~\rightarrow~1$ for $(r - R_x) >> a_x$, which 
defines the external region. Thus $R_x$ forms an approximate border 
between the internal and the external regions, and {\it if} $R_x$ is close 
to the radius of the nucleus and $a_x$ is of the order of 0.5~fm, then the 
two regions will correspond to the high density and low density regions 
of nuclear matter, respectively. This is indeed the case, as found in 
global fits to kaonic atom data \cite{MFG06}. The parameter $b_0$ 
represents the low-density interaction and the parameter $B_0$ represents 
the interaction inside the nucleus. We note that, unlike with pionic and 
antiprotonic atoms, the dependence of kaonic atom fits on the rms radius 
of the neutron distribution is weak, and the explicit inclusion of 
isovector terms, such as $b_1$ of Eq.~(\ref{eq:Vopt}), has only marginal 
effect.

\begin{figure}[t]
\centering
\includegraphics[height=8cm,width=7.8cm]{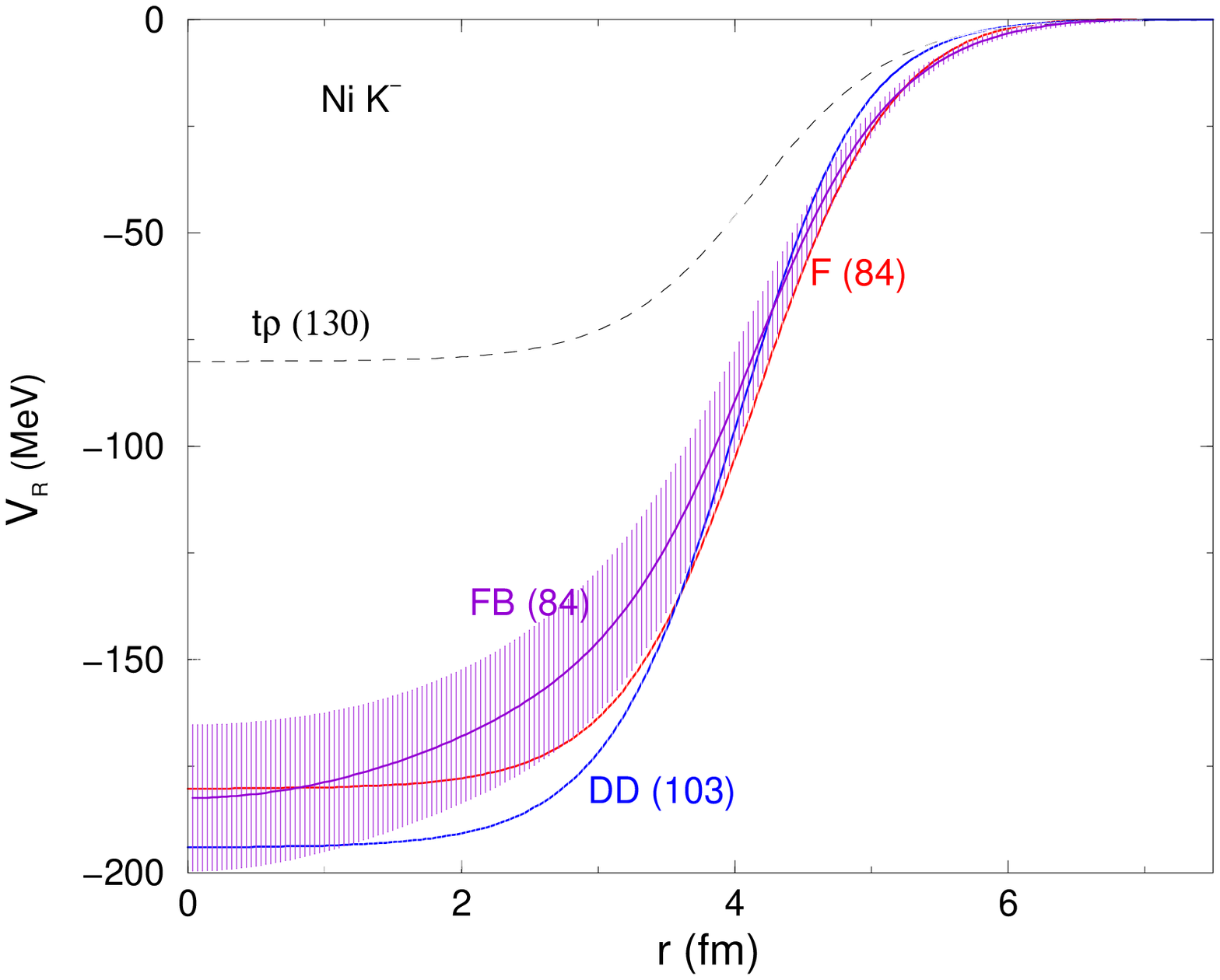}
\hspace*{3mm}
\includegraphics[height=8cm,width=8.0cm]{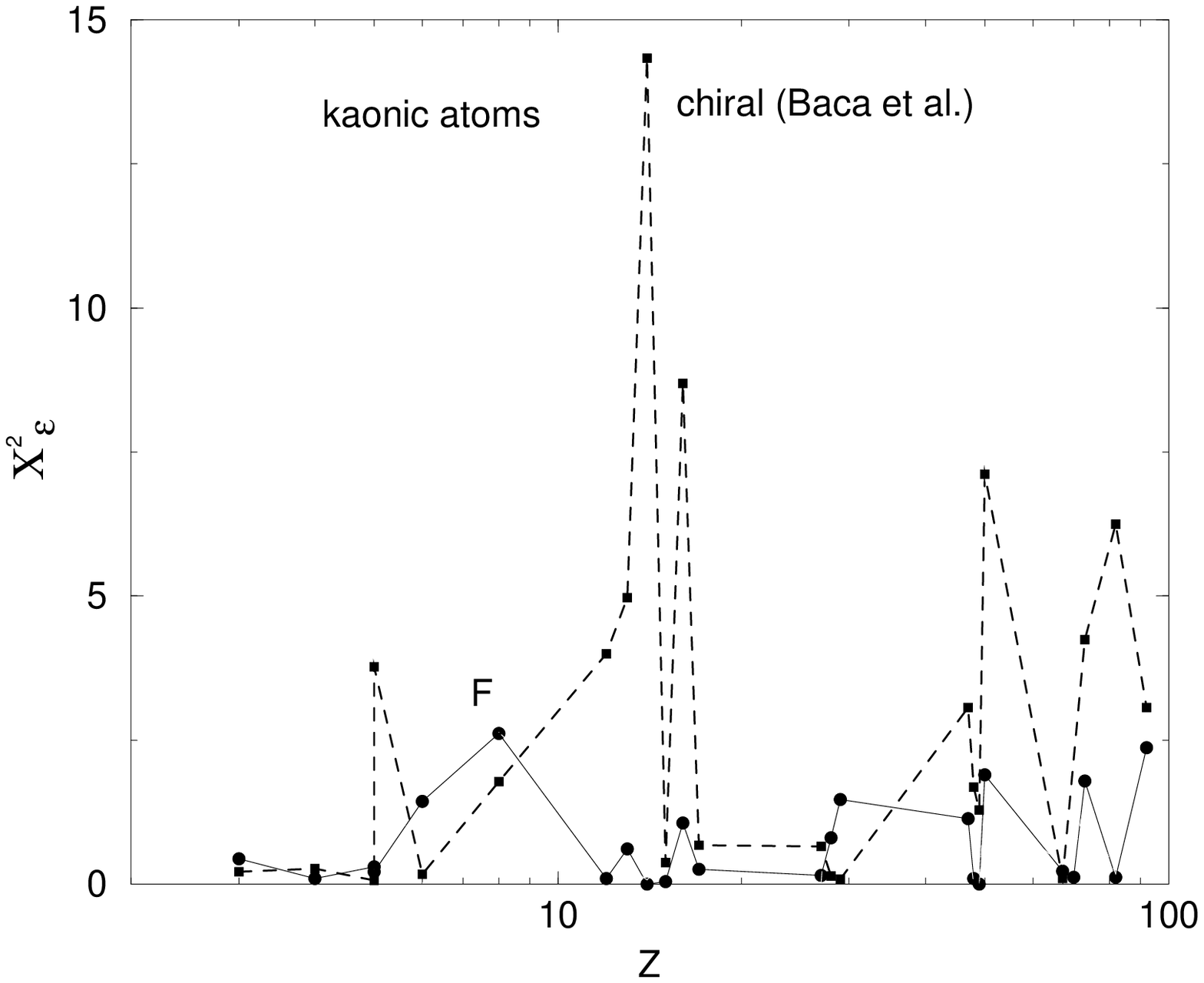} 
\caption{Left: real part of the ${\bar K}$-${^{58}}$Ni potential
obtained in a global fit to $K^-$-atom data using the model-independent
FB technique~{\protect \cite{BFr07}}, in comparison with
other best-fit potentials and  $\chi^2$ values in parentheses.
Right: contributions to the $\chi^2$ of $K^-$ atomic shifts for the {\it deep}
density-dependent potential F from Ref.~{\protect \cite{MFG06}} and for the
{\it shallow} chirally-based potential from Ref.~{\protect \cite{BGN00}}.} 
\label{fig:kbarNiVR}
\end{figure}

Figure \ref{fig:kbarNiVR} (left) shows, as an example, the real part
of the best-fit potential for $^{58}$Ni obtained with the various
models discussed above, i.e. the simple $t \rho $ model and its 
DD extension, and the geometrical model F, with the corresponding values of
$\chi ^2$ for 65 data points in parentheses. Also shown, with an error
band, is a Fourier-Bessel (FB) fit \cite{BFr07} that is discussed below.
We note that, although the two density-dependent
potentials marked DD and F have very different parameterizations, the
resulting potentials are quite similar. In particular, the shape of potential
F departs appreciably from  $\rho (r)$ for $\rho (r)/\rho_0 \leq 0.2$, where
the physics of the $\Lambda(1405)$ is expected to play a role.
The density dependence of the potential F provides by far the best fit ever 
reported for any global $K^-$-atom data fit, and the lowest $\chi ^2$ value 
as reached by the model-independent FB method.
On the right-hand side of the figure are shown the individual contributions
to $\chi ^2$ of the shifts for the deep F potential and the shallow 
chirally-based potential (of depth about 50 MeV) due to Baca et 
al. \cite{BGN00}. It is self evident that the agreement between calculation 
and experiment is substantially better for the deep F potential than for 
the shallow chiral potential. 

The question of how well the real part of the
$K^-$-nucleus potential is determined was discussed in Ref.~\cite{BFr07}.
Estimating the uncertainties of hadron-nucleus potentials as function
of position is not a simple task. For example, in the `$t\rho $'
approach the shape of the potential is determined by the nuclear
density distribution and the uncertainty in the strength parameter,
as obtained from $\chi ^2$ fits to the data,
implies a fixed {\it relative} uncertainty at all radii, which is, of course,
unfounded. Details vary when more elaborate
forms such as DD or F are used, but one is left essentially
with {\it analytical continuation} into the nuclear interior of potentials
that might be well-determined only close to the nuclear surface.
`Model-independent' methods have been used in analyses of elastic scattering
data for various projectiles \cite{BFG89}
to alleviate this problem. However, applying e.g. the Fourier-Bessel (FB) 
method in global analyses of kaonic atom data, one ends up
with too few terms in the series, thus making the uncertainties unrealistic
in their dependence on position.
This is illustrated in Fig. \ref{fig:kbarNiVR} by the FB curve, 
obtained by adding a Fourier-Bessel series to a $t\rho $ potential. 
Only three terms in the series are needed to achieve a $\chi ^2$ of 84
and the potential becomes deep, in agreement with the other two `deep'
solutions. The error band obtained from the FB method \cite{BFG89} is, 
nevertheless, unrealistic because only three FB terms are used. However, 
an increase in the number of terms is found to be unjustified numerically.

\begin{figure}[t]
\centerline{\includegraphics[height=9cm]{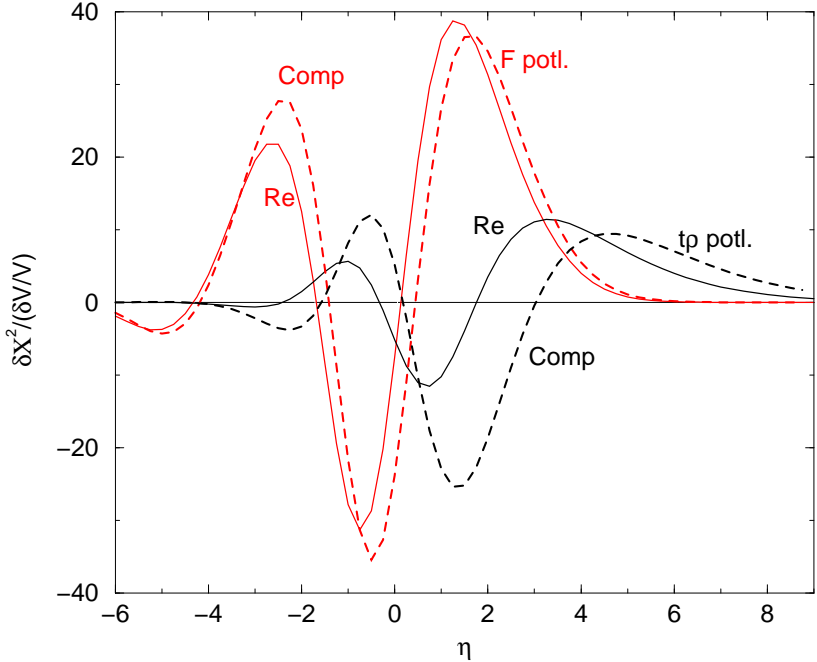}}
\caption{Functional derivatives of kaonic atoms $\chi ^2$ with respect to 
the fully complex (Comp, dashed lines) and real (Re, solid lines) potential 
as function of $\eta$, where $r=R_c+\eta a_c$, with $R_c$ and $a_c$ the 
radius and diffuseness parameters, respectively, of a 2pF charge 
distribution. Results are shown for the $t\rho $ and for the F potentials 
of Ref.~\cite{MFG06} obtained from global fits to kaonic atom data.} 
\label{fig:katFD}
\end{figure}

The functional derivative (FD) method for identifying the radial regions
to which exotic atom data are sensitive was described in detail in
Sect.~\ref{sec:fd}. This method was applied 
in Ref.~\cite{BFr07} to the F and $t \rho $ kaonic atom
potentials and results are shown in Fig.~\ref{fig:katFD}
where $\eta$ is a global parameter defined by
$r=R_c+\eta a_c$, with $R_c$ and $a_c$ the radius and diffuseness
parameters, respectively, of a 2pF charge distribution.
From the figure it can be inferred that the sensitive region
for the real $t\rho $ potential is between $\eta =-1.5$ and $\eta =6$
whereas for the F potential it is between $\eta =-3.5$ and $\eta =4$.
Recall that $\eta =-2.2$ corresponds to 90\% of the central charge density
and $\eta =2.2$ corresponds to 10\% of that density. It therefore becomes 
clear that within the $t\rho $ potential there is no sensitivity to the 
interior of the nucleus whereas with the density-dependent F potential,
which yields greatly improved fit to the data, there is sensitivity
to regions within the full nuclear density.
The different sensitivities result from
the potentials themselves: for the $t \rho $ 
potential the interior of the nucleus
is masked essentially by the strength of the imaginary potential.
In contrast, for the F potential not only is its imaginary part significantly
smaller than the imaginary part of the $t \rho $ potential 
\cite{MFG06} but also
the additional attraction provided
by the deeper potential enhances the {\it atomic} wavefunctions within
the nucleus \cite{BFG97} thus creating the sensitivity at smaller radii.
As seen in the figure, the functional derivative for the complex 
F potential is well approximated by that for its real part. 

The optical potentials derived from the observed strong-interaction 
effects in kaonic atoms are sufficiently deep to support deeply-bound 
antikaon {\it nuclear} states, but it does not necessarily imply that 
such states are sufficiently narrow to be resolved unambiguously from 
experimental spectra. Moreover, choosing between the very shallow 
chirally motivated potentials \cite{ROs00,CFG01}, the intermediate 
chiral potentials of depth around 100 MeV~\cite{WKW96} or the deep 
phenomenological potentials of type F adds appreciable ambiguity 
to predictions made for such states. 
It should also be kept in mind that these depths relate to $\bar K$ 
potentials at {\it threshold}, whereas the information required for 
$\bar K$-nuclear quasibound states is at energies of order 100 MeV below 
threshold. Predictions become model independent only when it comes to 
`deeply-bound' $K^-$ {\it atomic} states, as discussed below. 

\subsection{Deeply bound $K^-$ atomic states}
\label{sec:Kdeep}

\begin{figure}[t]
\centerline{\includegraphics[height=9cm]{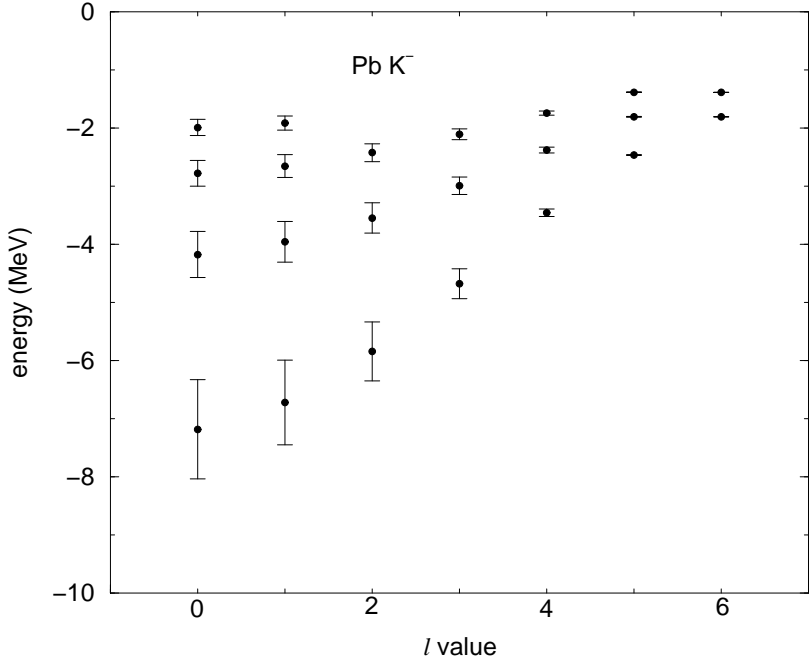}}
\caption{Calculated energies of $K^-$ atomic states in $^{208}$Pb.
The lowest energy for each $l$ value corresponds to $n=l+1$.
The bars represent the widths of the states.}
\label{fig:Kspect}
\end{figure}

Somewhat paradoxically, due to the strong absorptive imaginary part 
of the $K^-$-nucleus potential, relatively narrow deeply-bound 
atomic states are expected to exist which are quite independent of 
the real potential. Such states are indeed found in 
numerical calculations as demonstrated in  Fig.~\ref{fig:Kspect}
where calculated binding energies and widths of atomic 
states of $K^-$ in $^{208}$Pb are shown for several $l$-values, down
to states which are inaccessible via the X-ray cascade. For $^{208}$Pb, 
the last observed atomic circular state is the $7i$, corresponding 
to $l=6$. The general physics behind this phenomenon is similar to that
responsible for the deeply-bound pionic atom states, although there
are differences in the underlying mechanisms. 
The mechanism behind the pionic atom deeply bound states is simply
the {\it repulsive} real part of the $s$-wave potential which expels 
the atomic wavefunction $\psi_{\rm atom}$ from the nucleus, thus reducing 
the overlap between $\psi_{\rm atom}$ and the imaginary potential. 
This reduction, according to 
\begin{equation} 
\label{eq:gamma} 
\Gamma=-2 \frac{\int \left|\psi_{\rm atom}\right|^2 {\rm Im} V_{\rm opt}
d{\bf r}}{\int \left|\psi_{\rm atom}\right|^2 d{\bf r}}\;, 
\end{equation} 
results in a reduced width for atomic states. Eq.~(\ref{eq:gamma}) 
holds {\it exactly} for a Schr\"odinger equation, with only small changes 
for a KG equation, see Refs.~\cite{FGa99a,FGa99b}.  
In contrast, phenomenological kaonic atom potentials are {\it attractive}, 
but the strengths of the imaginary potential are such that the decay of 
$\psi_{\rm atom}$ as it enters the nucleus is equivalent to repulsion, 
resulting in narrow atomic states due to the reduced overlap as discussed 
above. It is seen from Fig.~\ref{fig:Kspect} that there is a saturation 
phenomenon where widths hardly increase for $l \leq 2$, contrary to 
intuitive expectation. The repulsive effect of sufficiently strong absorption 
is responsible for the general property of saturation of widths of atomic 
states and also for saturation of reaction cross sections above threshold, 
observed experimentally for antiprotons \cite{BFG01}. 

\begin{figure}[t]
\centering
\includegraphics[height=8.0cm,width=7.8cm]{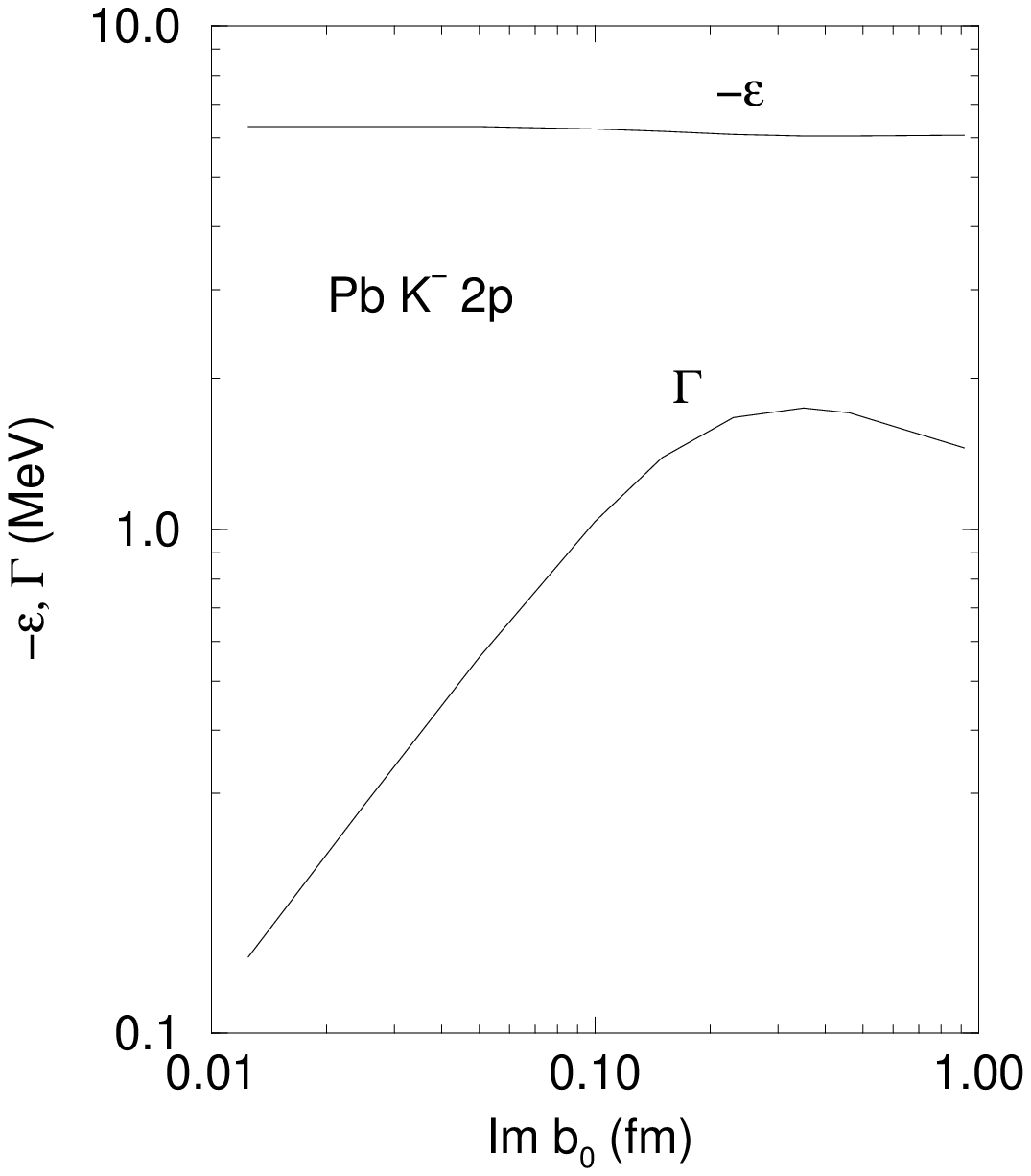}
\hspace*{3mm}
\includegraphics[height=8.0cm,width=7.8cm]{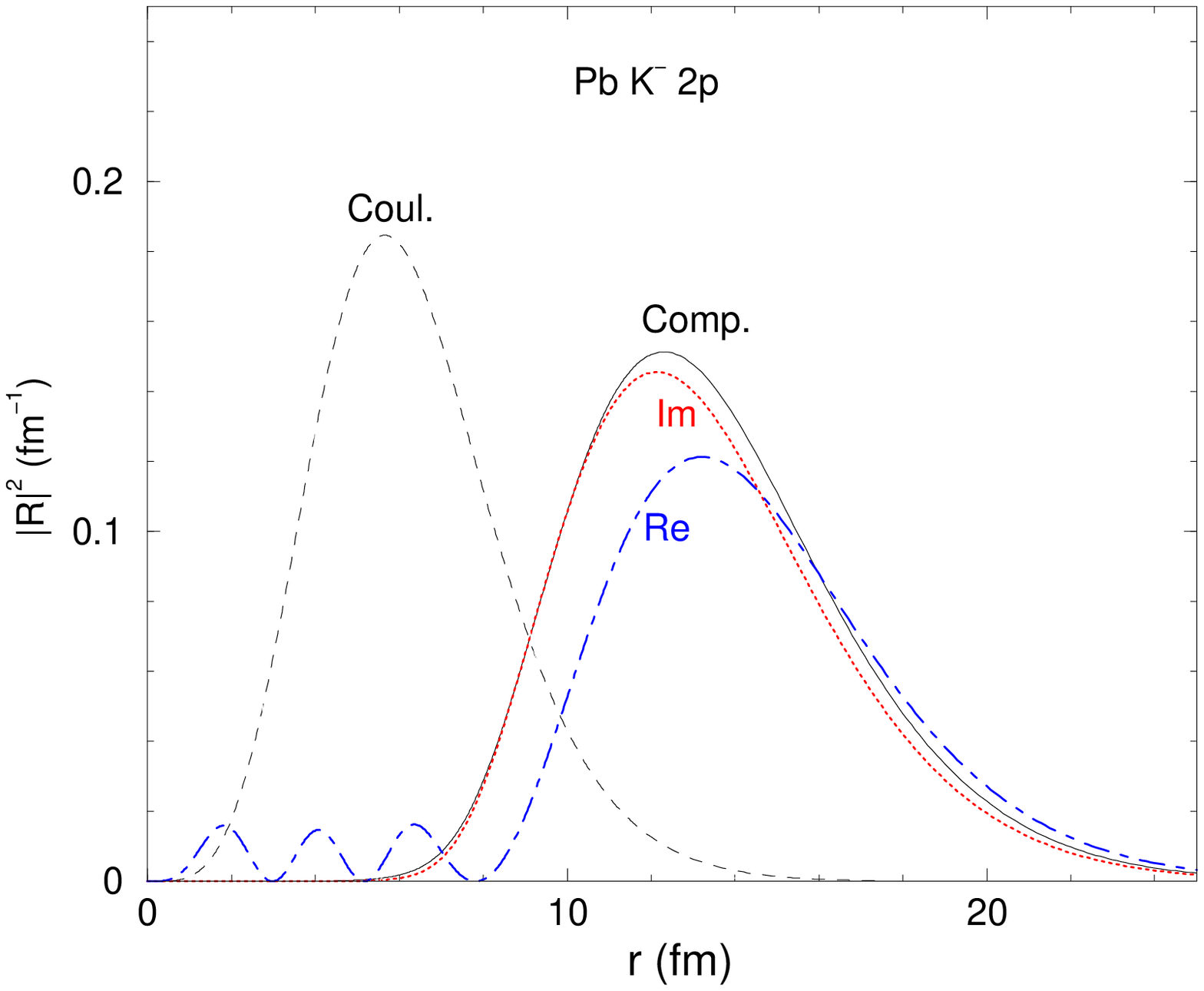}
\caption{Left: saturation of width $\Gamma$ for the $2p$ `deeply bound' 
$K^-$ atomic state in $^{208}$Pb as function of absorptivity Im~$b_0$, for
Re~$b_0 = 0.62$~fm. Right: wavefunctions for this state, see text.}
\label{fig:sat}
\end{figure}

The left-hand side of Fig.~\ref{fig:sat} shows the saturation of widths
as function of the absorptive strength parameter Im~$b_0$ of $V_{\rm opt}$, 
Eq.~(\ref{eq:Vopt}), for the 2$p$ state of kaonic atoms of $^{208}$Pb. 
For small values of Im~$b_0$ the calculated width increases linearly, 
but already at 20\% of the best-fit value of 0.9~fm saturation sets in 
and eventually the width goes down with further increase of the absorption. 
Note that the real part of the binding energy, represented here by the 
strong-interaction level shift $\epsilon$, hardly changes with Im~$b_0$. 
The right-hand side of Fig.~\ref{fig:sat} shows radial wavefunctions 
for the 2$p$  atomic $K^-$ state in $^{208}$Pb for several combinations 
of potentials. The dashed curve marked `Coul' is for the Coulomb potential 
only, and with a half-density radius for $^{208}$Pb of 6.7 fm it clearly 
overlaps strongly with the nucleus. Adding the full complex optical 
potential the solid curve marked `Comp' shows that the atomic 
wavefunction is expelled from the nucleus, and the dotted curve
marked `Im' shows that this repulsion is effected by the imaginary
part of the potential. Clearly the overlap of the atomic wavefunction 
with the nucleus is dramatically reduced compared to the Coulomb-only 
situation. An interesting phenomenon is displayed by the dot-dashed 
curve marked `Re'. It shows the atomic wavefunction when the real 
potential is added to the Coulomb potential, demonstrating significant 
{\it repulsion} of the atomic wavefunction by the added {\it attractive} 
potential. The explanation for this bizarre result is provided by the 
three small peaks inside the nucleus which are due to the orthogonality 
of the {\it atomic} wavefunction and strongly-bound $K^-$ {\it nuclear} 
wavefunctions having the same $l$-values. This extra structure of the 
atomic wavefunction in the interior effectively disappears when the 
imaginary potential is included.

\section{$\bf {\bar K}$ Nuclear Interactions} 
\label{sec:kstrong}  

\subsection{The $K^- p$ interaction near threshold} 
\label{sec:Kpnearth} 

\begin{figure}[t] 
\includegraphics[scale=0.7]{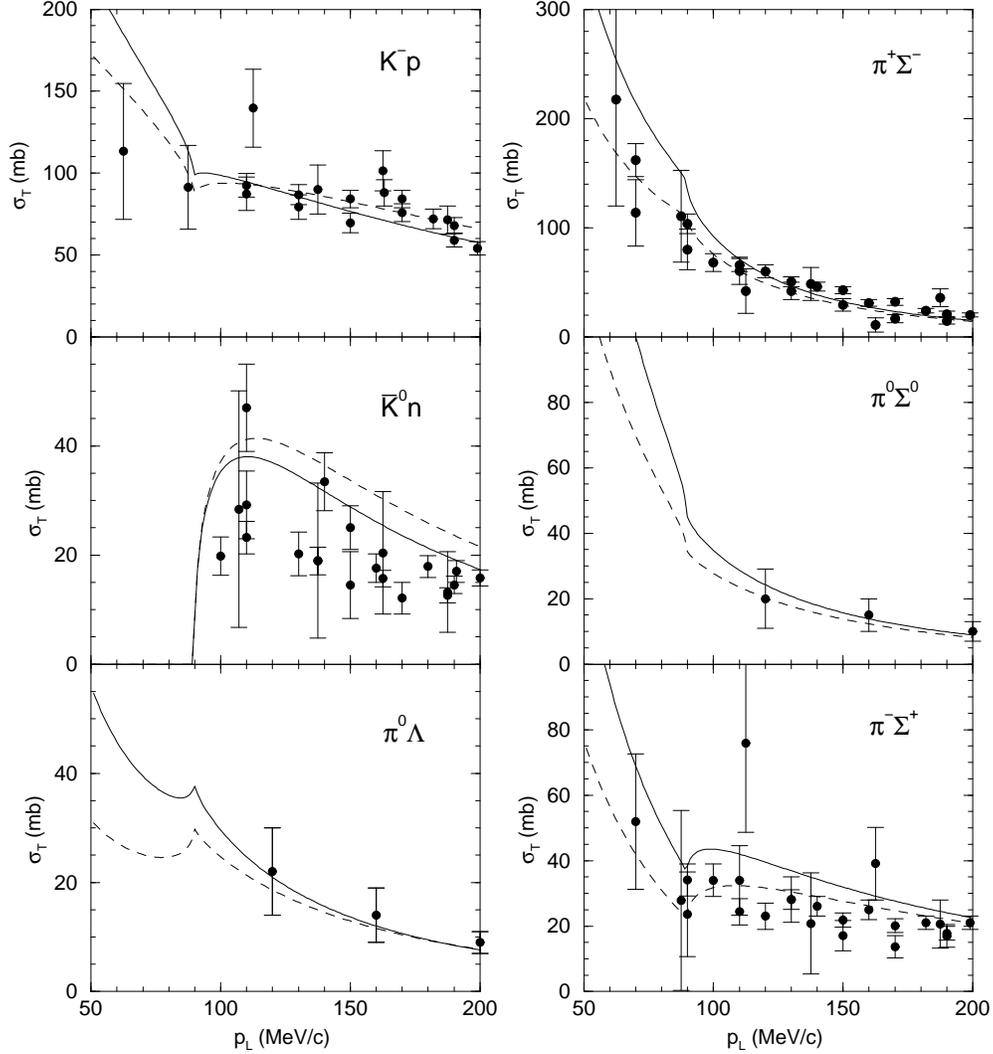} 
\caption{Calculations from Ref.~{\protect \cite{CFG01}} of cross sections
for $K^{-}p$ scattering and reactions. The dashed lines show free-space 
chiral-model coupled-channel calculations. The solid lines show  
chiral-model coupled-channel calculations using slightly varied parameters 
in order to fit also the $K^-$-atom data for a (shallow) optical potential 
calculated self consistently.}  
\label{fig:kminuspdata} 
\end{figure} 

The $K^- p$ data at low energies provide a good experimental base upon 
which models for the strong interactions of the $\bar K N$ system have 
been developed. Near threshold the coupling to the open $\pi \Sigma$ and 
$\pi \Lambda$ channels is extremely important, as may be judged from the 
size of the $K^- p$ reaction cross sections, particularly 
$K^- p \to \pi^+ \Sigma^-$, with respect to the $K^- p$ elastic cross 
sections shown in Fig.~\ref{fig:kminuspdata}. By developing potential models, 
$\bar K N$ amplitudes are obtained that allow for analytic continuation into 
the nonphysical region below $K^- p$ threshold. Using a K-matrix analysis, 
this was the way Dalitz and Tuan predicted the existence of the 
$\Lambda(1405)$ $\pi \Sigma$, $I=0$ resonance in 1959~\cite{DTu59}. 

\begin{figure}[t] 
\centerline{\includegraphics[height=7cm,width=9cm]{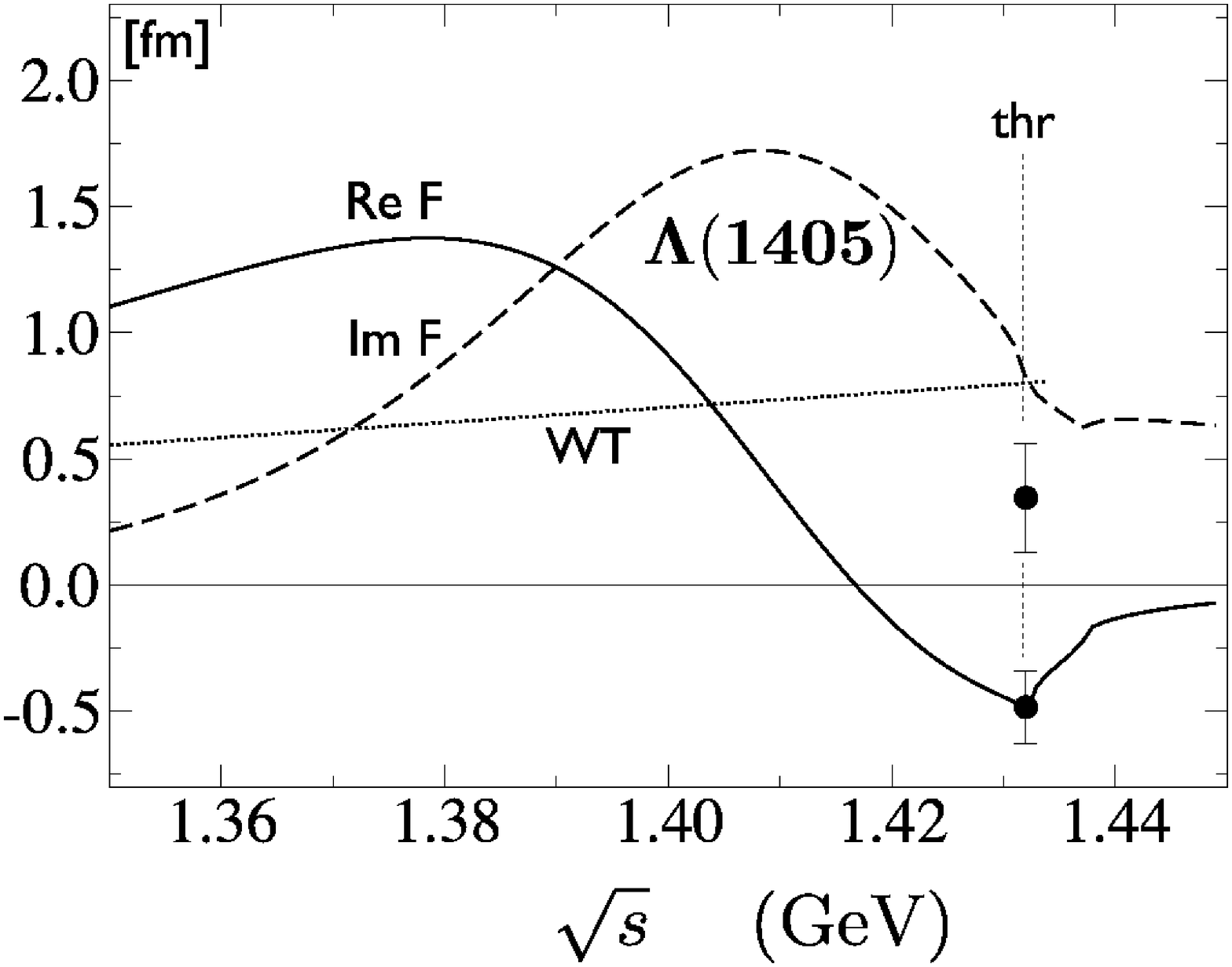}} 
\caption{Real and imaginary parts of the $K^- p$ forward elastic 
scattering amplitude, fitted within a NLO chiral SU(3) coupled-channel 
approach to $K^- p$ scattering and reaction data. The line denoted WT 
is the (real) LO Tomozawa-Weinberg $K^- p$ driving-term amplitude. 
The DEAR measurement~{\protect \cite{BBC05}} value for $a_{K^- p}$ is 
shown with error bars. Figure taken from Ref.~{\protect \cite{Wei07}}, 
based on the work of Ref.~{\protect \cite{BNW05b}}.} 
\label{fig:weisehyp06} 
\end{figure}

A recent example from coupled-channel potential model calculations 
\cite{BNW05a,BNW05b,BNW06,BMN06}, based on low-energy chiral expansion 
of meson-baryon potentials in the $S=-1$ sector, is shown in 
Fig.~\ref{fig:weisehyp06} where the real and imaginary parts of the 
resulting $K^- p$ elastic scattering amplitude, continued analytically 
below the $K^- p$ threshold, are plotted. The line marked WT stands for 
the leading Weinberg-Tomozawa (WT) nonresonant $K^- p$ amplitude below 
threshold when channel-coupling effects are switched off. The figure 
demonstrates that the $\Lambda(1405)$ resonance is generated 
{\it dynamically} within the coupled-channel calculation. A discrepancy 
with Im~$a_{K^-p}$ deduced from the DEAR measurement~\cite{BBC05} is 
highlighted in this figure. In contrast, the purely $I=1$ $K^- n$ amplitude 
does not show such resonance effects below threshold, and its chiral model 
dependence is considerably weaker than the model dependence of amplitudes 
affected by the $\Lambda(1405)$ resonance, e.g. the $K^- p$ elastic 
scattering amplitude shown in Fig.~\ref{fig:weisehyp06}.  

\subsection{$\bar K$-nucleus potentials} 
\label{sec:kbarpot} 

The gross features of low-energy $\bar K N$ physics, as demonstrated 
in the previous section by chiral coupled-channel fits to the low-energy 
$K^-p$ scattering and reaction data, are encapsulated in the lowest-order 
(LO) WT vector term of the chiral effective Lagrangian~\cite{WRW97}. 
The Born approximation for the $\bar K$-{\it nuclear} optical potential 
$V_{\bar K}$ due to the driving-term WT interaction yields then a sizable 
attraction: 
\begin{equation} 
\label{eq:chiral} 
V_{\bar K}=-\frac{3}{8f_{\pi}^2}~\rho\sim -55~\frac{\rho}{\rho_0}~~{\rm MeV} 
\end{equation} 
for $\rho _0 = 0.16$ fm$^{-3}$, where $f_{\pi} \sim 93$ MeV is the 
pseudoscalar meson decay constant. Iterating the TW term plus 
next-to-leading-order (NLO) terms, 
within an {\it in-medium} coupled-channel approach constrained 
by the $\bar K N - \pi \Sigma - \pi \Lambda$ data near the 
$\bar K N$ threshold, roughly doubles this $\bar K$-nucleus attraction 
as may be seen by inspecting Fig.~\ref{fig:weisehyp06}. 
It is found (e.g. Ref.~\cite{WKW96}) that the $\Lambda(1405)$ quickly 
dissolves in the nuclear medium at low density, so that 
the repulsive free-space scattering length $a_{K^-p}$, as function of 
$\rho$, becomes {\it attractive} well below $\rho _0$. Since the purely 
$I=1$ attractive scattering length $a_{K^-n}$ is only weakly density 
dependent, the resulting in-medium $\bar K N$ isoscalar scattering length 
$b_0(\rho)={\frac{1}{2}}(a_{K^-p}(\rho)+a_{K^-n}(\rho)$) translates into 
a strongly attractive $V_{\bar K}$: 
\begin{equation} 
\label{eq:trho} 
V_{\bar K}(r) = -{\frac{2\pi}{\mu_{KN}}}~b_0(\rho)~\rho(r)~, 
~~~~{\rm Re}V_{\bar K}(\rho_0) \sim -110~{\rm MeV}\,. 
\end{equation} 
However, when $V_{\bar K}$ is calculated {\it self consistently}, 
including $V_{\bar K}$ in the propagator $G_0$ used in the 
Lippmann-Schwinger equation determining $b_0(\rho)$, one obtains 
Re$V_{\bar K}(\rho_0)\sim -$(40-60) MeV~\cite{SKE00,ROs00,TRP01,CFG01}. 
The main reason for this weakening of $V_{\bar K}$, 
approximately going back to that calculated using Eq.~(\ref{eq:chiral}), 
is the strong absorptive effect which $V_{\bar K}$ exerts within $G_0$ to 
suppress the higher Born terms of the $\bar K N$ TW potential. 

Additional considerations for estimating  $V_{\bar K}$ are listed below.  
 
\begin{itemize} 

\item QCD sum-rule estimates~\cite{Dru06} for vector (v) and scalar (s) 
self-energies: 
\begin{eqnarray} 
\label{eq:QCDv} 
\Sigma_v(\bar K) &\sim & -\frac{1}{2}~\Sigma_v(N)~\sim~
-\frac{1}{2}~(200)~{\rm MeV}~ =~-100~{\rm MeV}\,,\\ 
\Sigma_s(\bar K) &\sim & \frac{m_s}{M_N}~\Sigma_s(N)~\sim~
\frac{1}{10}~(-300)~{\rm MeV}~ =~ -30~{\rm MeV}\, ,
\label{eq:QCDs}
\end{eqnarray}
where $m_s$ is the strange-quark (current) mass. The factor 1/2 in 
Eq.~(\ref{eq:QCDv}) is due to the one nonstrange antiquark $\bar q$ in the 
$\bar K$ meson out of two possible, and the minus sign is due to G-parity 
going from $q$ to $\bar q$. This rough estimate gives then 
$V_{\bar K}(\rho_0) \sim -130$~MeV. 

\item The QCD sum-rule approach essentially 
refines the mean-field argument~\cite{SGM94,BRh96} 
\begin{equation} 
\label{eq:MF} 
V_{\bar K}(\rho_0)~\sim~\frac{1}{3}~(\Sigma_s(N)-\Sigma_v(N))~\sim~
-170~{\rm MeV}\,,
\end{equation} 
where the factor 1/3 is again due to the one nonstrange antiquark in the 
$\bar K$ meson, but here with respect to the three nonstrange quarks of 
the nucleon. 

\item The ratio of $K^-/K^+$ production cross sections in nucleus-nucleus and
proton-nucleus collisions near threshold, measured by the Kaon Spectrometer 
(KaoS) collaboration~\cite{SBD06} at SIS, GSI, yields an estimate 
$V_{\bar K}(\rho_0) \sim -80$~MeV by relying on BUU transport calculations 
normalized to the value $V_K(\rho_0) \sim +25$~MeV. 
Since $\bar K NN \to YN$ absorption processes 
apparently were disregarded in these calculations, a deeper $V_{\bar K}$ may 
follow once nonmesonic absorption processes are included. 

\end{itemize}

\subsection{Deeply bound $K^-$ nuclear states in light nuclei}
\label{sec:nuclear db}

The first prediction of a $\bar K$-nuclear quasibound state was made by 
Nogami~\cite{Nog63} as early as 1963, arguing that the $I=1/2,~L=S=0$ state 
of the $K^-pp$ system could be bound by about 10 MeV. Recent calculations 
confirm this prediction, with higher values of binding energies but also with 
substantial values for the (mesonic) width of this state, as summarized in 
Table~\ref{tab:kpp}. We note that the Faddeev calculations listed in the 
table account rigorously for the strong $I=0~\bar K N \to \Sigma N$ coupling, 
but all the calculations overlook the $\bar K NN \to YN$ coupling to 
nonmesonic channels which are estimated to add conservatively 20 MeV to 
the overall width. If $K^-pp$, the lightest possible $\bar K$-nuclear 
system, is indeed bound, then it is plausible that heavier $\bar K$-nuclear 
systems will also possess quasibound states and the remaining question is 
whether these states are sufficiently narrow to allow observation and 
identification. Unlike the saturation of width in $K^-$ atoms, discussed in 
Sect.~\ref{sec:Kdeep}, no saturation mechanism holds for the width of 
$\bar K$-nuclear states which retain very good overlap with the potential.  

\begin{table}
\caption{Binding energies ($B$) and widths ($\Gamma$) calculated for 
$K^-pp$ (in MeV).} 
\label{tab:kpp} 
\begin{ruledtabular} 
\begin{tabular}{lcccc} 
channels &\multicolumn{2}{c}{single channel} &\multicolumn{2}{c}
{coupled channels} \\ 
Ref. & ATMS~\cite{YAk02} & AMD~\cite{DWe06} & Faddeev~\cite{SGM06,SGM07}& 
Faddeev~\cite{ISa06,ISa07} \\ \hline 
$B$  & 48 & 20--50 & 50--70 & 60--95 \\ 
$\Gamma$ & 61 & --  & 90--110 & 45--80 \\ 
\end{tabular} 
\end{ruledtabular} 
\end{table} 

Ongoing experiments by the FINUDA spectrometer collaboration at 
DA$\Phi$NE, Frascati, already claimed evidence for a relatively 
broad $K^- pp$ deeply bound state ($B \sim 115$~MeV) by observing 
back-to-back $\Lambda p$ pairs from the decay $K^-pp\to\Lambda p$ 
in $K^{-}_{\rm stop}$ reactions on Li and $^{12}$C~\cite{ABB05}, 
but these pairs could naturally arise from conventional absorption 
processes at rest when final-state interaction is taken into 
account~\cite{MOR06}. 
Indeed, the $K^-_{\rm stop}pn\to \Sigma^- p$ reaction observed recently 
in $^6$Li~\cite{ABB06} does not require any $K^-d$ quasibound state. 
It is worth noting, however, that in order to search for a $K^-pn$ bound 
state which is charge symmetric to the $K^-pp$ quasibound state discussed 
above, one should use a $^7$Li target to look for back-to-back $\Sigma^- p$ 
pairs. Very recently, a $\Lambda p$ narrow peak has been reported in 
$\bar p$ annihilation on $^4$He from the OBELIX spectrometer data at LEAR, 
CERN \cite{BBF07}, corresponding to a yet deeper $K^-pp$ quasibound state 
($B \sim 160$ MeV) if this interpretation is valid, given the reservations 
mentioned above. A definitive study of the $K^- pp$ quasibound state 
(or more generally $\{\bar K[NN]_{I=1}\}_{I=1/2}$) could be reached 
through fully exclusive formation reactions, such as:
\begin{equation} 
\label{eq:nag}
K^-+^3{\rm He}~ \to ~ n + \{\bar K[NN]_{I=1}\}_{I=1/2,I_z=+1/2},~~~ 
p + \{\bar K[NN]_{I=1}\}_{I=1/2,I_z=-1/2} \, , 
\end{equation} 
the first of which is scheduled for day-one experiment in J-PARC~\cite{Nag06}. 
We note that the large widths calculated for the $K^-pp$ quasibound state 
could make it difficult to identify the state experimentally~\cite{KHa07}. 

The current experimental and theoretical interest in $\bar K$-nuclear bound 
states was triggered back in 1999 by the suggestion of Kishimoto~\cite{Kis99} 
to look for such states in $(K^{-},p)$ reactions in flight, and by Akaishi and 
Yamazaki~\cite{AYa99,AYa02} who suggested to look for a $\bar K NNN$ $I=0$ 
state bound by over 100 MeV for which the main $\bar K N \to \pi \Sigma$ 
decay channel would be kinematically closed. In fact, Wycech had conjectured 
that the width of such states could be as small as 20 MeV~\cite{Wyc86}. 
Evidence claimed initially for relatively narrow states in the inclusive 
$(K^{-}_{\rm stop},p)$ and $(K^{-}_{\rm stop},n)$ spectra on $^4$He has 
recently been withdrawn \cite{Iwa06,SBC07a}, just to be replaced by 
a complementary low-statistics $\Lambda d$ narrow peak reported in 
$\bar p$ annihilation on $^4$He \cite{BBF07}, corresponding to 
a quasibound $\bar K NNN~I=0$ state with $B \sim 120$ MeV. 
Such correlated $\Lambda d$ pairs could arise from secondary 
three-nucleon absorption processes, as recently discussed by the FINUDA 
\cite{ABB07} and the KEK \cite{SBC07b} Collaborations 
in $K^{-}_{\rm stop}$ reactions on $^6$Li and $^4$He, respectively.  
On heavier targets, enhancements have been observed in the $(K^{-},n)$ 
in-flight spectrum on $^{16}$O \cite{KHA05}, but subsequent $(K^{-},n)$ 
and $(K^{-},p)$ reactions on $^{12}$C at $p_{\rm lab}=1$~GeV/c have not 
disclosed any peaks beyond the appreciable strength observed below the 
$\bar K$-nucleus threshold \cite{KHA07}. It is clear that the issue 
of $\bar K$ nuclear states is far yet from being experimentally resolved 
and more dedicated, systematic searches are necessary.

\subsection{RMF dynamical calculations of $\bar K$ quasibound nuclear states} 
\label{sec:RMF} 

\begin{figure}[t]
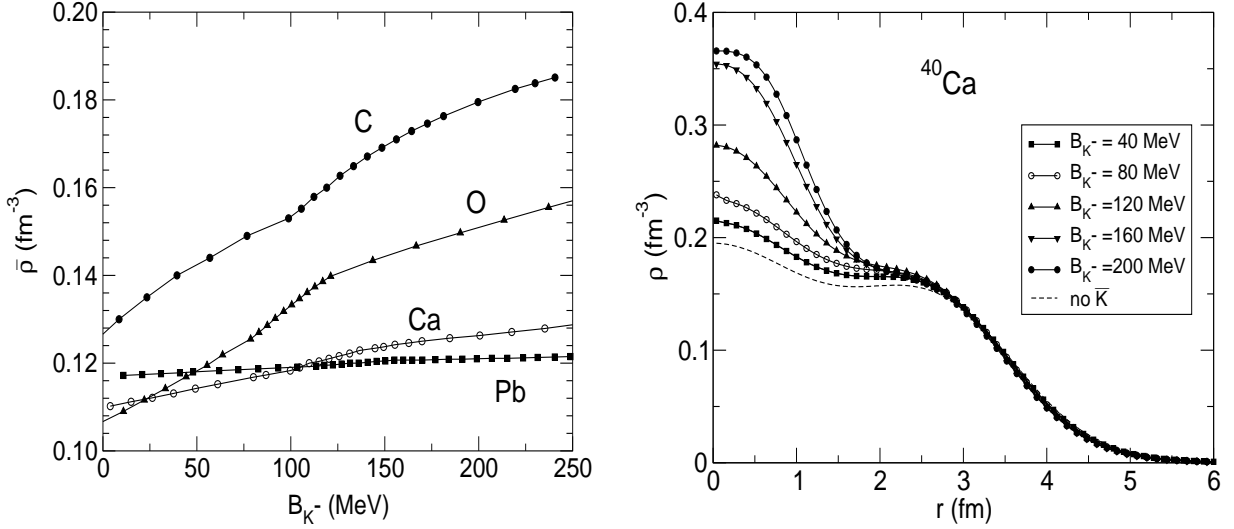
 
\centering 
\includegraphics[height=7cm,width=7.8cm]{K05fig6.eps} 
\hspace*{3mm} 
\includegraphics[height=7cm,width=7.8cm]{K05fig8.eps} 
\caption{Left: dynamically calculated average nuclear density $\bar \rho$ of 
$1s$ $K^-$-nuclear states in the nuclei denoted, as function of the $1s$ 
$K^-$ binding energy. Right: dynamically calculated nuclear density $\rho$ 
of $^{~~40}_{K^-}$Ca for several $1s$ $K^-$ nuclear states with specified 
$B_{K^-}$ values~\cite{MFG06}.}
\label{fig:rho} 
\end{figure}

In this model, spelled out in Refs.~\cite{MFG05,MFG06,GFG07}, the (anti)kaon 
interaction with the nuclear medium is incorporated by adding to ${\cal L}_N$ 
the Lagrangian density ${\cal L}_K$: 
\begin{equation}
\label{eq:Lk}
{\cal L}_{K} = {\cal D}_{\mu}^*{\bar K}{\cal D}^{\mu}K -
m^2_K {\bar K}K
- g_{\sigma K}m_K\sigma {\bar K}K\; .
\end{equation} 
The covariant derivative
${\cal D_\mu}=\partial_\mu + ig_{\omega K}{\omega}_{\mu}$ describes the 
coupling of the (anti)kaon to the vector meson $\omega$. The (anti)kaon 
coupling to the isovector $\rho$ meson was found to have negligible effects. 
The $\bar K$ meson induces additional source terms in the equations of motion 
for the meson fields $\sigma$ and $\omega_0$. It thus affects the scalar 
$S = g_{\sigma N}\sigma$ and the vector $V = g_{\omega N}\omega_0$ potentials 
which enter the Dirac equation for nucleons, and this leads to rearrangement 
or polarization of the nuclear core, as shown on the left-hand side of 
Fig.~\ref{fig:rho} for the calculated average nuclear density 
$\bar \rho = \frac{1}{A}\int\rho^2d{\bf r}$ 
as a function of $B_{K^-}$ for $K^-$ nuclear $1s$ states across the periodic 
table, and on the right-hand side of the figure for the density of 
$^{~~40}_{K^-}$Ca for several $1s$ $K^-$ nuclear states with specified 
$B_{K^-}$ values~\cite{MFG06}. It is seen that in the light $K^-$ nuclei, 
$\bar \rho$ increases substantially with $B_{K^-}$ to values about 50\% 
higher than without the $\bar K$. The increase of the central nuclear 
densities is bigger, up to 50-100\%, and is nonnegligible even in the 
heavier $K^-$ nuclei where it is confined to a small region of order 1.5~fm. 
Furthermore, in the Klein-Gordon 
equation satisfied by the $\bar K$, the scalar $S = g_{\sigma K}\sigma$ and 
the vector $V = -g_{\omega K}\omega_0$ potentials become 
{\it state dependent} through the {\it dynamical} density dependence of the 
mean-field potentials $S$ and $V$, as expected in a RMF calculation. 
An imaginary ${\rm Im}V_{\bar K} \sim t\rho$ was added, fitted to the 
$K^-$ atomic data~\cite{FGM99}. It was then suppressed by an energy-dependent 
factor $f(B_{\bar K})$, considering the reduced phase-space for the initial 
decaying state and assuming two-body final-state kinematics for the decay 
products in the $\bar K N \to \pi Y$ mesonic modes ($80\%$) and in the 
$\bar K NN \to Y N$ nonmesonic modes ($20\%$).  

\begin{figure}
\includegraphics[scale=0.5]{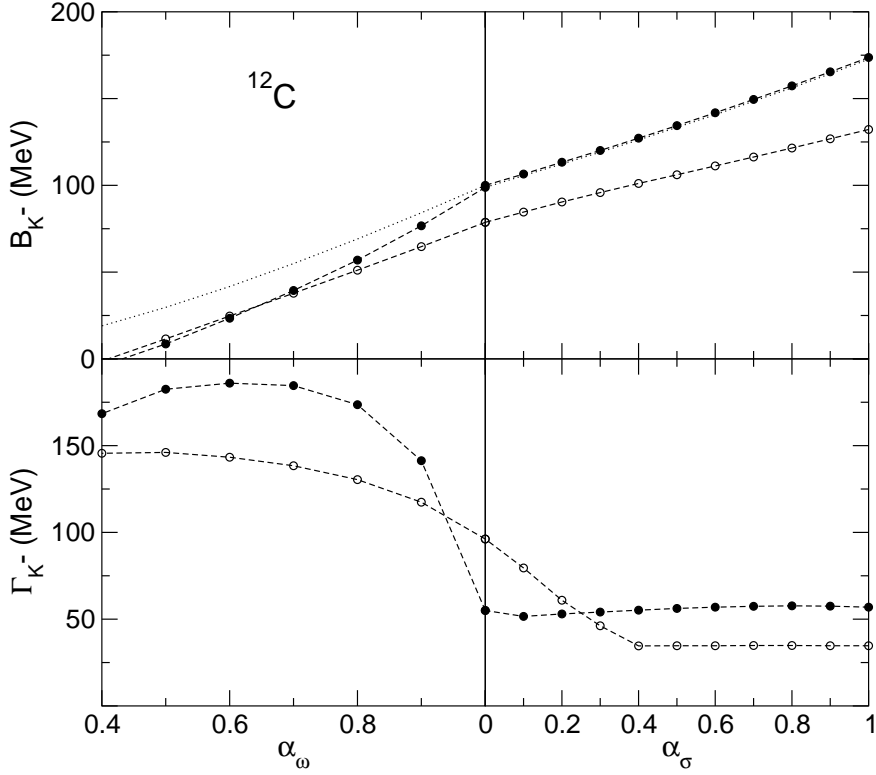} 
\caption{$1s$ $K^-$ binding energy and width in $^{~~12}_{K^-}$C calculated 
statically (open circles) and dynamically (solid circles) for the nonlinear 
RMF model NL-SH {\protect \cite{SNR93}} as function of the $\omega K$ and 
$\sigma K$ coupling strengths: $\alpha_{\omega}$ is varied in the left panels 
as indicated, with $\alpha_{\sigma}=0$, and $\alpha_{\sigma}$ is varied in 
the right panels as indicated, with $\alpha_{\omega}=1$. The dotted line 
shows the calculated binding energy when the absorptive $K^-$ potential is 
switched off in the dynamical calculation.}
\label{fig:dynam}
\end{figure}

The RMF coupled equations were solved self-consistently. For a rough idea, 
whereas the static calculation gave $B_{K^-}^{1s}=132$~MeV
for the $K^-$ $1s$ state in $^{12}$C, using the values 
$g^{\rm atom}_{\omega K},~g^{\rm atom}_{\sigma K}$ from the 
$K^-$-atom fit, the dynamical calculation gave $B_{K^-}^{1s}=172$~MeV. 
In order to scan a range of values for $B_{K^-}^{1s}$, the coupling 
constants $g_{\sigma K}$ and $g_{\omega K}$ were varied in given intervals 
of physical interest. An example is shown in Fig.~\ref{fig:dynam}. 

Beginning approximately with $^{12}$C, the following conclusions may be drawn: 

\begin{itemize} 

\item For given values of $g_{\sigma K},g_{\omega K}$, the $\bar K$ binding 
energy $B_{\bar K}$ saturates as function of $A$, except for a small increase 
due to the Coulomb energy (for $K^-$). 

\item The difference between the binding energies calculated dynamically and 
statically, $B_{\bar K}^{\rm dyn} - B_{\bar K}^{\rm stat}$, is substantial 
in light nuclei, increasing with $B_{\bar K}$ for a given value of $A$, 
as shown in the upper panels of Fig.~\ref{fig:dynam}, and 
decreasing monotonically with $A$ for a given value of $B_{\bar K}$. 
It may be neglected only for very heavy nuclei. The same holds for the 
nuclear rearrangement energy $B_{\bar K}^{\rm s.p.} - B_{\bar K}$ which is 
a fraction of $B_{\bar K}^{\rm dyn} - B_{\bar K}^{\rm stat}$. 
 
\item The functional dependence of the width $\Gamma_{K^-}(B_{K^-})$, 
shown for $^{~~12}_{K^-}$C in the lower panels of Fig.~\ref{fig:dynam} 
follows the shape of the suppression factor $f(B_{K^-})$ which falls off 
rapidly until $B_{K^-} \sim 100$~MeV, where the dominant
$\bar K N \rightarrow \pi \Sigma$ gets switched off, and then stays
rather flat in the range $B_{K^-} \sim$~100-200~MeV where the width is 
dominated by the $\bar K NN \to YN$ absorption modes. The widths
calculated dynamically in this range are considerably larger than if 
calculated statically. Adding the residual width neglected in this calculation, 
due to the $\bar K N \to \pi \Lambda$ secondary mesonic decay channel, and 
assigning these two-nucleon absorption modes a $\rho^2$ density dependence, 
a lower limit of $\Gamma_{\bar K} \gtrsim 50$~MeV is obtained for 
deeply-bound states in the range $B_{K^-} \sim$~100-200~MeV \cite{GFG07}. 

\end{itemize}

\subsection{Kaon condensation} 
\label{sec:cond} 

The possibility of kaon condensation in dense matter was proposed by 
Kaplan and Nelson \cite{KNe86,NKa87}, with subsequent works offering 
related scenarios in nuclear matter \cite{BLR94,LBM94}. Neutron stars, 
with a density range extending to several times nuclear-matter density, 
have been considered extensively as the most natural dense systems 
where kaon condensation is likely to be realized. It is commonly 
accepted that under some optimal conditions, kaon condensation could 
occur at densities above $3\rho_0$ depending on the way hyperons 
enter the constituency of neutron stars. However, our concern here is 
not with neutron stars where time scales of the weak interactions are 
operative, enabling the transformation $n \to p + K^-$ or a rare weak 
decay such as $e^- \to K^- + \nu_e$ to transform `high-energy' electrons 
to antikaons once the effective mass of $K^-$ mesons dropped below 
200 MeV approximately. Our concern here is limited to laboratory 
strong-interaction processes where hadronization and equilibration 
time scales in collisions leading to dense matter are much shorter, 
of order fm/c. If antikaons bind strongly to nuclei, 
then one might ask whether or not the binding energy 
per $\bar{K}$ meson in multi-$\bar{K}$ nuclear states increases 
significantly upon adding a large number of $\bar{K}$ mesons, 
so that $\bar{K}$ mesons provide the physical degrees of freedom for 
self-bound strange hadronic systems. Precursor phenomena to kaon 
condensation in nuclear matter would occur beyond some threshold value 
of strangeness, if the binding energy $B_{\bar{K}}$ per $\bar{K}$ meson 
exceeds the combination $m_Kc^2 + \mu_N - m_{\Sigma}c^2 \gtrsim 240$~MeV, 
where $\mu_N$ is the nucleon chemical potential. Furthermore, once 
$B_{\bar{K}} \gtrsim m_Kc^2 + \mu_N - m_{\Lambda}c^2 \gtrsim 320$~MeV, 
$\Lambda$, $\Sigma$ and $\Xi$ hyperons would no longer combine 
macroscopically with nucleons to compose the more conventional kaon-free 
form of strange hadronic matter \cite{BGa00}. 

\begin{figure}[t]
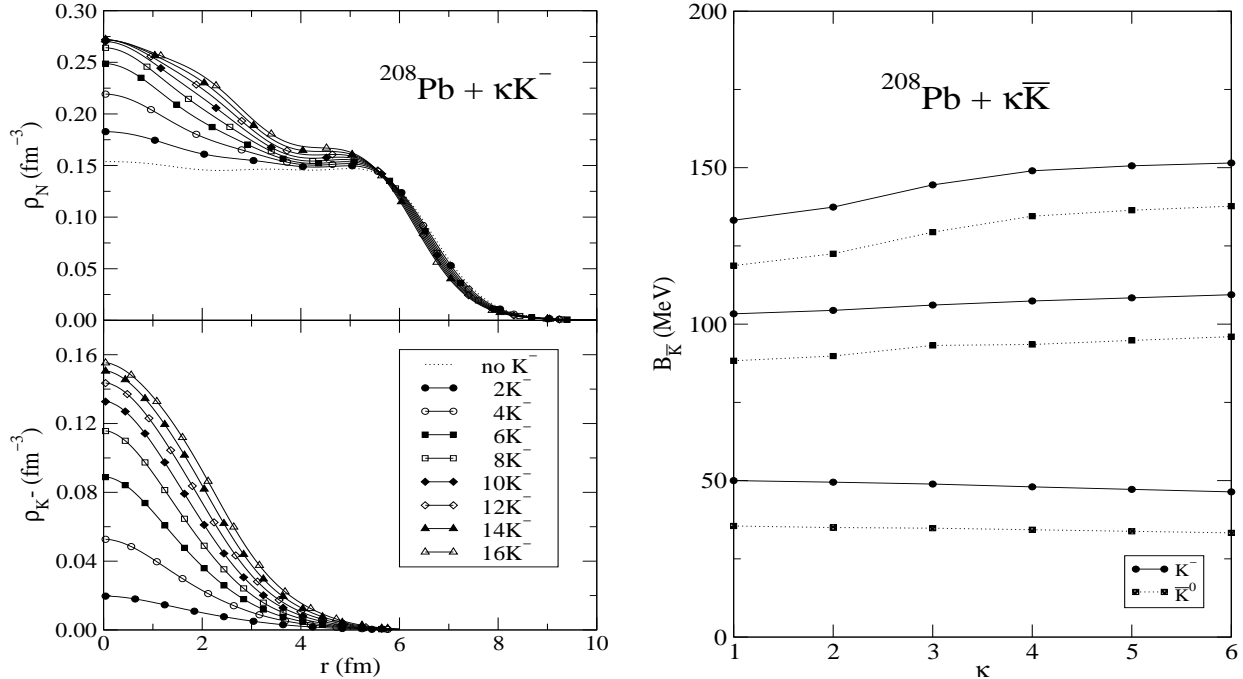

\centering
\includegraphics[height=9cm,width=8.0cm]{gfgm07fig11.eps} 
\hspace*{3mm}
\includegraphics[height=9cm,width=7.8cm]{gfgm07fig9.eps} 
\caption{Left: nuclear density $\rho_N$ (top panel) and $1s$ ${\bar K}$ 
density $\rho_{\bar K}$ (bottom panel) in $^{208}{\rm Pb} + \kappa K^-$, 
starting with $B_{K^-}=100$~MeV in $^{208}{\rm Pb} + 1K^-$. The dotted 
curve stands for the $^{208}{\rm Pb}$ density in the absence of $\bar K$ 
mesons. Right: 1s $\bar K$ binding energy $B_{\bar K}$ in $^{208}{\rm Pb} 
+ \kappa {\bar K}$, where ${\bar K} = K^-$ (circles) and ${\bar K}^0$ 
(squares). Figure taken from Ref.~{\protect \cite{GFG07}}.} 
\label{fig:multi}  
\end{figure} 

Gazda et al.~\cite{GFG07} recently calculated multi-$\bar{K}$ nuclear 
configurations, finding that the nuclear and $\bar K$ densities behave 
regularly upon increasing the number of antikaons embedded in the 
nuclear medium, without any indication for abrupt or substantial increase 
of the densities. The central nuclear densities appear to saturate at 
approximately 50\% higher values than the central nuclear density with 
one antikaon, as shown on the left-hand side of Fig.~\ref{fig:multi} 
for multi-$K^-$ $^{208}{\rm Pb}$ nuclei. Furthermore, the $\bar K$ 
binding energy saturates upon increasing the number of $\bar K$ mesons 
embedded in the nuclear medium. The heavier the nucleus is, the more 
antikaons it takes to saturate the binding energies, but even for 
$^{208}{\rm Pb}$ the number required does not exceed approximately 10, 
as shown on the right-hand side of Fig.~\ref{fig:multi}. We note that 
the interaction between antikaons in this extended RMF calculation is 
mediated by isoscalar boson fields: vector $\omega$ and $\phi$, and scalar 
$\sigma$. The binding-energy saturation owes its robustness to the 
dominance of the {\it repulsive} vector interactions over the attractive 
scalar interactions for antikaon pairs. The saturated values of $\bar K$ 
binding energies do not exceed the range of values 100--200 MeV considered 
normally as providing deep binding for one antikaon. This range of binding 
energies leaves antikaons in multi-${\bar K}$ nuclei comfortably above the 
range of energies where hyperons might be relevant. It is therefore 
unlikely that multi-${\bar K}$ nuclei may offer precursor phenomena in 
nuclear matter towards kaon condensation.

\section{${\bf \Sigma}$ Hyperons}
\label{sec:sigma}

\subsection{Overview} 
\label{sec:overv} 
 
One Boson Exchange (OBE) models fitted to the scarce low-energy $YN$ 
scattering data produce within a $G$-matrix approach, with one exception 
(Nijmegen Model F), as much attraction 
for the $\Sigma$ nuclear potential as they 
do for the $\Lambda$ nuclear potential, see Ref.~\cite{DGa84} for a review 
of `old' models and Ref.~\cite{RYa06} for the latest state of the art for 
Nijmegen models. Indeed, the best-fit $t_{\rm eff}\rho$ potential for 
$\Sigma^-$ atoms was found by Batty et al.~\cite{Bat79,BGT83} to be attractive 
and absorptive, with central depths for the real and imaginary parts of 
25-30~MeV and 10-15~MeV, respectively. It took almost a full decade, 
searching for $\Sigma$ hypernuclear bound states at CERN, KEK and BNL, 
before it was realized that except for a special case for $^4_\Sigma$He, 
the observed continuum $\Sigma$ hypernuclear spectra indicate a very shallow, 
or even repulsive $\Sigma$ nuclear potential, 
as reviewed by Dover et al.~\cite{DMG89}. 
These indications have received firm support with the measurement of several 
$(K^-,\pi^{\pm})$ spectra at BNL \cite{BCF99} followed by calculations for 
$^9$Be~\cite{Dab99}. Recently, with measurements of the $\Sigma^-$ spectrum 
in the $(\pi^-,K^+)$ reaction taken at KEK across the periodic table 
\cite{NSA02,SNA04}, it has become established that the $\Sigma$ nuclear 
interaction is strongly repulsive. In parallel, analyses of $\Sigma^-$-atom 
in the early 1990s, allowing for density dependence or 
departure from the $t \rho$ prescription, motivated mostly by the precise 
data for W and Pb \cite{PEG93}, led to the conclusion 
that the {\it nuclear} interaction of $\Sigma$s is dominated by 
repulsion~\cite{BFG94a,BFG94b,MFG95}, as reviewed in Ref.~\cite{BFG97}. 
This might have interesting repercussions for the balance of 
strangeness in the inner crust of neutron stars~\cite{BGa97}, primarily by 
delaying the appearance of $\Sigma^-$ hyperons to higher densities, if at all, 
as discussed below. The inability of the Nijmegen OBE models, augmented by 
G-matrix calculations~\cite{RYa06}, to produce $\Sigma$ nuclear repulsion 
is a serious drawback for these models at present. This problem apparently 
persists also in the Juelich model approach~\cite{HMe05}. The only theoretical 
works that provide exception are SU(6) quark-model RGM calculations by the 
Kyoto-Niigata group~\cite{KFF00}, in which a strong Pauli repulsion appears in 
the $I=3/2,~{^3S_1}-{^3D_1}~\Sigma N$ channel, and Kaiser's SU(3) chiral 
perturbation calculation~\cite{Kai05} which yields repulsion of order 60 MeV 
in nuclear matter. 

\begin{figure}[t] 
\includegraphics[scale=0.7]{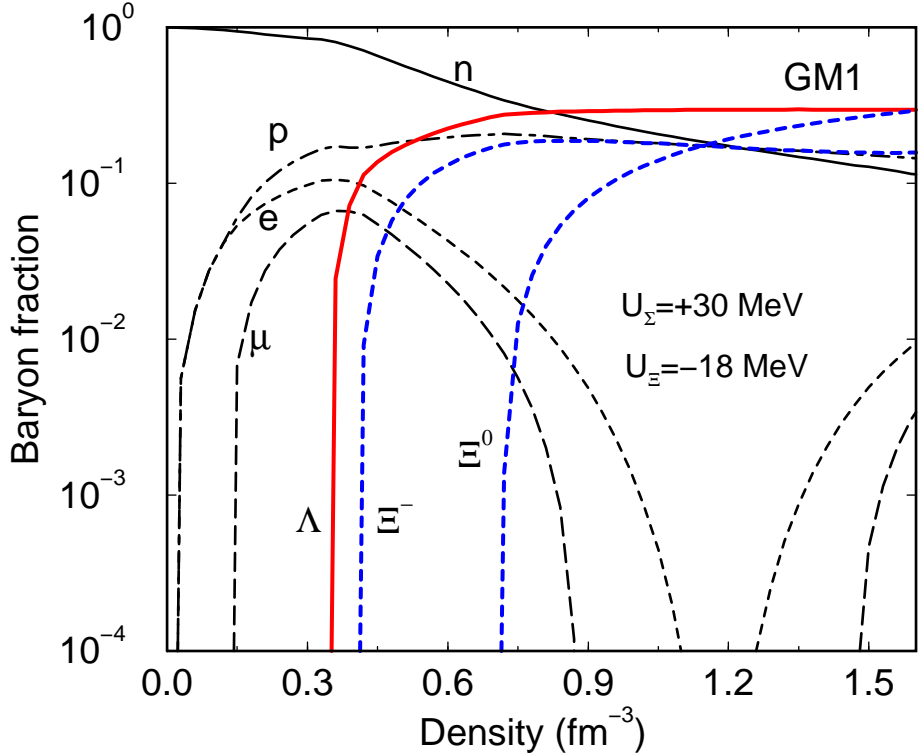} 
\caption{Fractions of baryons and leptons in neutron-star matter for a RMF 
calculation using set GM1 with weak $YY$ interactions 
{\protect \cite{SMi96}}. Figure taken from Ref.~{\protect \cite{SBi07}}. } 
\label{fig:nstars} 
\end{figure}

Since $\Sigma^-$ is the first hyperon (as function of density) to appear 
in neutron stars when the hyperon interactions are disregarded, it is 
natural to expect that the composition of neutron-star matter depends 
sensitively on the $\Sigma^-$-hypernuclear potential. For attractive 
$\Sigma^-$-hypernuclear potentials of the order of 30 MeV depth, as for 
$\Lambda$ hyperons in $\Lambda$ hypernuclei, the $\Sigma^-$ is indeed 
the first hyperon to appear, at density lower than twice nuclear matter 
density. However, for a repulsive potential, the situation reverses 
dramatically as shown in Fig.~\ref{fig:nstars}. Incidentally, a $K^-$ 
condensed phase might then appear at density between 3 to 4 times 
nuclear matter density replacing $\Xi^-$ hyperons and second in 
strangeness only to $\Lambda$ hyperons \cite{RSW01}. 

Below we briefly review and update the $\Sigma^-$ atom fits and the recent 
$(\pi^-,K^+)$ KEK results and their analysis. 

\subsection{Fits to $\Sigma^-$ atoms} 
\label{sec:SigDD} 

\begin{figure}[t] 
\includegraphics[scale=0.7]{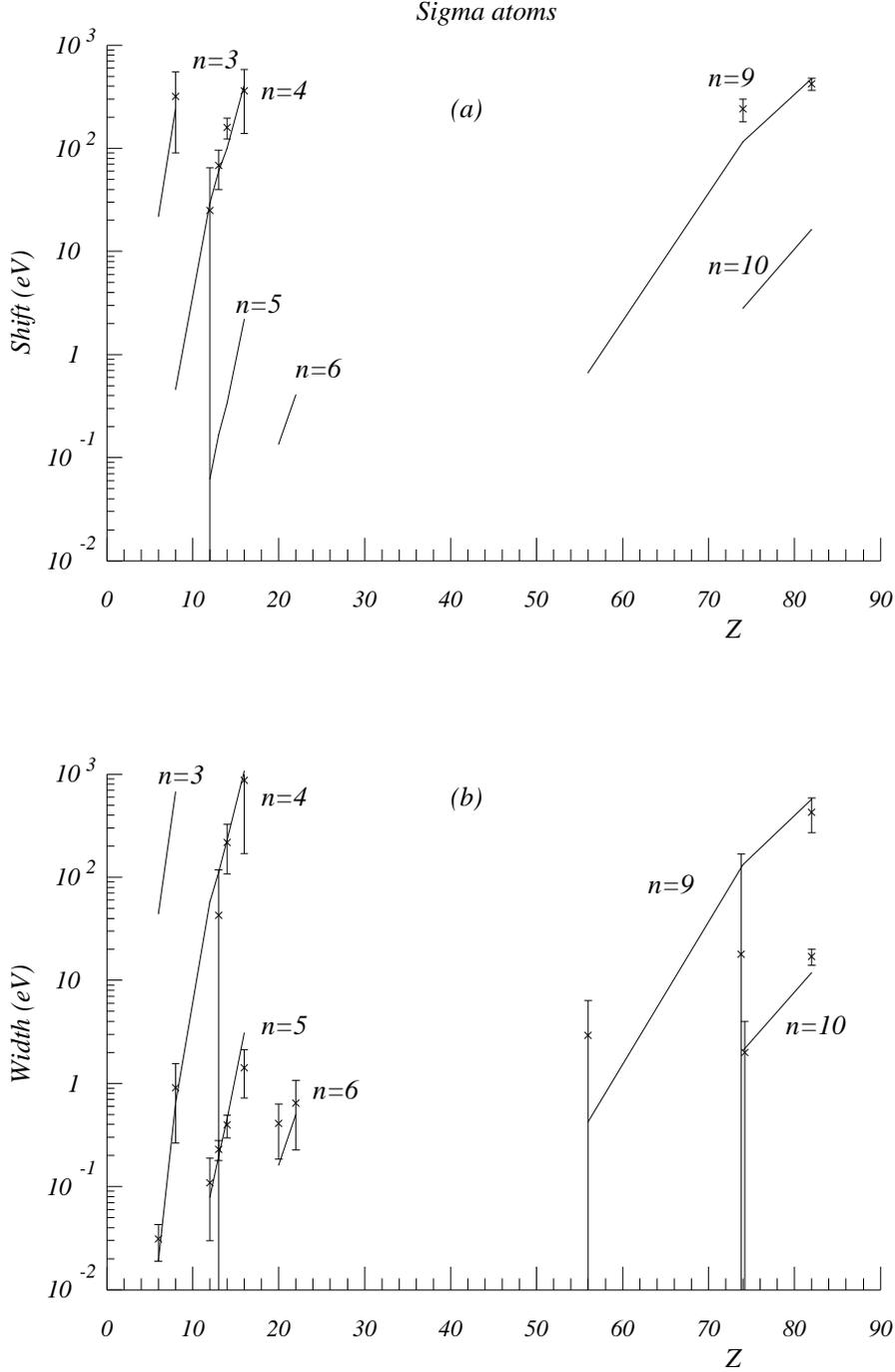} 
\caption{Shift and width values for sigma atoms. The continuous lines 
join points calculated with a best-fit DD optical potential, see next 
figure.} 
\label{fig:sigatoms} 
\end{figure} 

\begin{figure}
\includegraphics[scale=0.7]{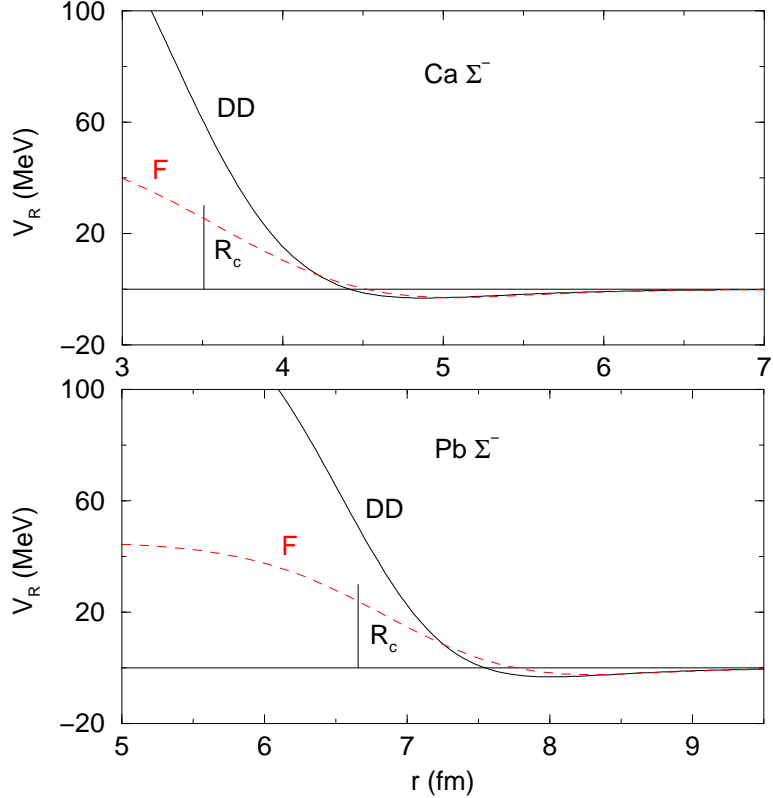} 
\caption{Re$V_{\rm opt}$  for DD (solid) and for the geometrical
model F (dashed)
$\Sigma^-$ nuclear potentials fitted to $\Sigma^-$ atomic 
data. 
Vertical bars indicate the half-density radius of 
the nuclear charge distribution.}
\label{fig:ddsig} 
\end{figure}

The data used in the $\Sigma^-$-atom fits are shown in Fig.~\ref{fig:sigatoms} 
representing all published measurements from C to Pb inclusive. The data are 
relatively inaccurate, reflecting the difficulty in making measurements of 
strong-interaction effects in $\Sigma^-$ atoms where most of the X-ray lines 
are relatively weak and must be resolved from the nuch stronger $K^-$ atomic 
X-ray transitions. Batty et al.~\cite{BFG94a,BFG94b} analyzed the full data 
set of $\Sigma^-$ atoms, consisting of strong-interaction level shifts, 
widths and yields, introducing a phenomenological density dependent (DD) 
potential of the isoscalar form

\begin{equation}
\label{eq:dd}
V_\Sigma(r) \sim \left[ b_0 + B_0
\left({\rho(r)/\rho(0)}\right)^\alpha \right]
\rho(r) \quad , \qquad \alpha > 0 \quad ,
\end{equation}
and fitting the parameters $b_0, B_0$ and $\alpha$ to the
data, greatly improved fits to the data are obtained. 
Isovector components are readily included in Eq.~(\ref{eq:dd}) but
are found to have a marginal effect. Note, however, that the
absorption was assumed to take place only on protons. 
The complex parameter $b_0$ may be identified with the
spin-averaged $\Sigma^-N$ scattering length. For the
best-fit isoscalar potentials, Re$V_\Sigma$ is attractive
at low densities outside the nucleus, changing into
repulsion in the nuclear surface region.
The precise magnitude and shape of the repulsive
component within the nucleus is not determined by the
atomic data. The resulting potentials are shown in Fig.~\ref{fig:ddsig} 
(DD, solid lines),
where it is worth noting that the transition from attraction to 
repulsion occurs well outside of the nuclear radius, hence the 
occurrence of this transition should be largely model independent. 
To check this last point we have repeated the fits to the atomic data
with the `geometrical model' F of Sect.~\ref{sec:Kat}, using separate
$t \rho $ expressions in an internal and an external region, see
Eq.~(\ref{eq:DDF}). The neutron densities used in the fits were of the 
skin type, with the $r_n-r_p$ parameter Eq.~(\ref{eq:RMF}) $\gamma$=1.0~fm. 
The fits deteriorate significantly if the halo type is used for the neutron
density. The fit to the data is equally good  with this model as with 
the DD model, ($\chi ^2$ per degree of freedom of 0.9 here 
compared to 1.0 for the DD model)
and the potentials are shown as the dashed lines in Fig.~\ref{fig:ddsig}.
The half-density radius of the charge distribution is indicated in
the figure. It is clear that both models show weak attraction at
large radii, turning into repulsion approximately one fm outside of 
that radius.

\begin{figure}
\includegraphics[scale=0.7]{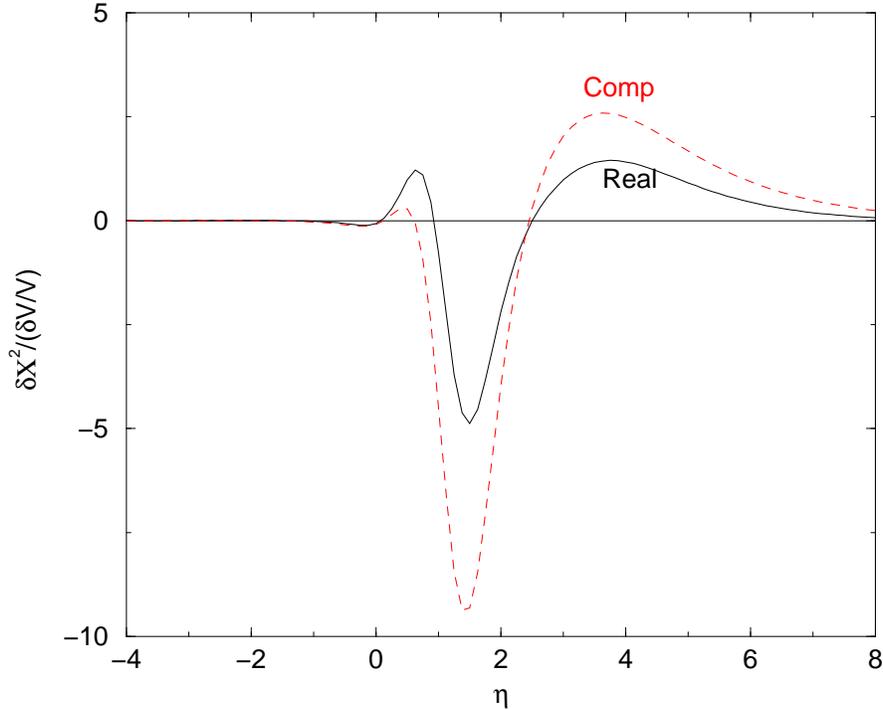}
\caption{Functional derivatives of $\chi ^2$ with respect to the real
(solid) and with respect to the full complex (dashed) optical potentials
for the best-fit F potential.}
\label{fig:SigFD}
\end{figure}

Further insight into the geometry of the $\Sigma$-nucleus interaction
is gained by inspecting the functional derivatives (FD) of $\chi ^2$
with respect to the optical potentials, see Sect.~\ref{sec:fd}. 
Figure~\ref{fig:SigFD} shows the FDs based on the best fit of 
the geometrical model F as
discussed above. From the differences between the FD with respect to the
full complex potential and the FD with respect to the real potential
it is concluded that both real and imaginary parts play similar roles
in the $\Sigma$-nucleus interaction.
The bulk of $|$FD$|$ is in the range of 0.5~$\leq~\eta~\leq$~6, 
covering the radial region where the weak attraction turns into repulsion. 
Obviously no information is obtained from $\Sigma^-$ atoms on the interaction
inside the nucleus.
It is also interesting to note quite generally that such 
 potentials do not produce bound states, and this
conclusion is in agreement with the experimental results from BNL 
\cite{BCF99} for the absence of $\Sigma$ hypernuclear peaks beyond He. 

Some semi-theoretical support for this finding of inner repulsion is given 
by RMF calculations by Mare{\v s} et al.~\cite{MFG95} who generated the 
$\Sigma$-nucleus interaction potential in terms of scalar ($\sigma$) and 
vector ($\omega,\rho$) meson mean field contributions, fitting its coupling 
constants to the relatively accurate $\Sigma^-$ atom shift and width data 
in Si and in Pb. The obtained potential fits very well the whole body 
of data on $\Sigma^-$ atoms. This potential, which is generally attractive 
far outside the nucleus, becomes repulsive at the nuclear surface and 
remains so inward in most of the acceptable fits, of order 10-20 MeV. 
The Pb data~\cite{PEG93} are particularly important in pinning 
down the isovector component of the potential which 
in this model is sizable and which, 
for $\Sigma^-$, acts against nuclear binding in core nuclei with $N-Z > 0$, 
countering the attractive Coulomb interaction. 
 On the other hand, 
for very light nuclear cores and perhaps only for $A = 4$ hypernuclei, 
this isovector component (Lane term) generates binding of 
$\Sigma^+$ configurations. In summary, the more modern fits to $\Sigma^-$ 
atom data~\cite{BFG94a,BFG94b,MFG95} and the present
fits with the geometrical model support the presence of a substantial 
repulsive component in the $\Sigma$-nucleus potential which 
excludes normal $\Sigma$-nuclear binding, except perhaps in very 
special cases such as $^4_\Sigma$He \cite{Hay89,HSA90,Nag98,Har98}. 

\subsection{Evidence from $(\pi^-,K^+)$ spectra}
\label{sec:pi-K+spec} 

\begin{figure} 
\includegraphics[height=14cm,width=9cm]{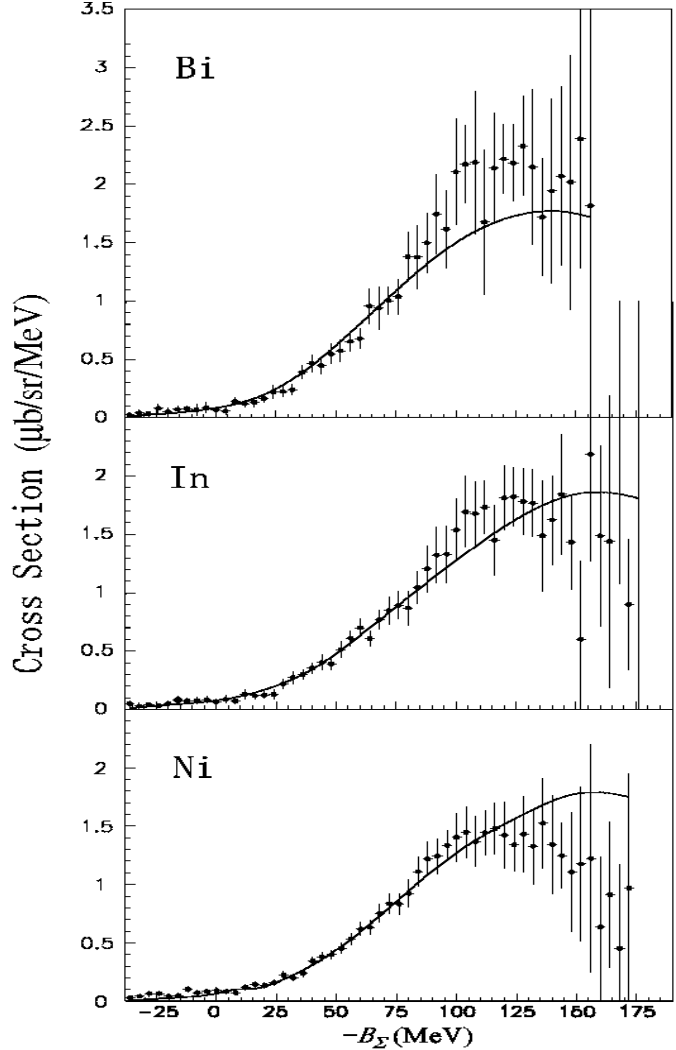} 
\caption{Inclusive $(\pi^-,K^+)$ spectra on Ni, In and Bi, fitted by 
a $\Sigma$-nucleus WS potential with depths $V_0 = 90$ MeV, $W_0=-40$ MeV 
{\protect\cite{SNA04}}.}
\label{fig:ninbispec} 
\end{figure}

\begin{figure}
\includegraphics[scale=0.8,angle=0]{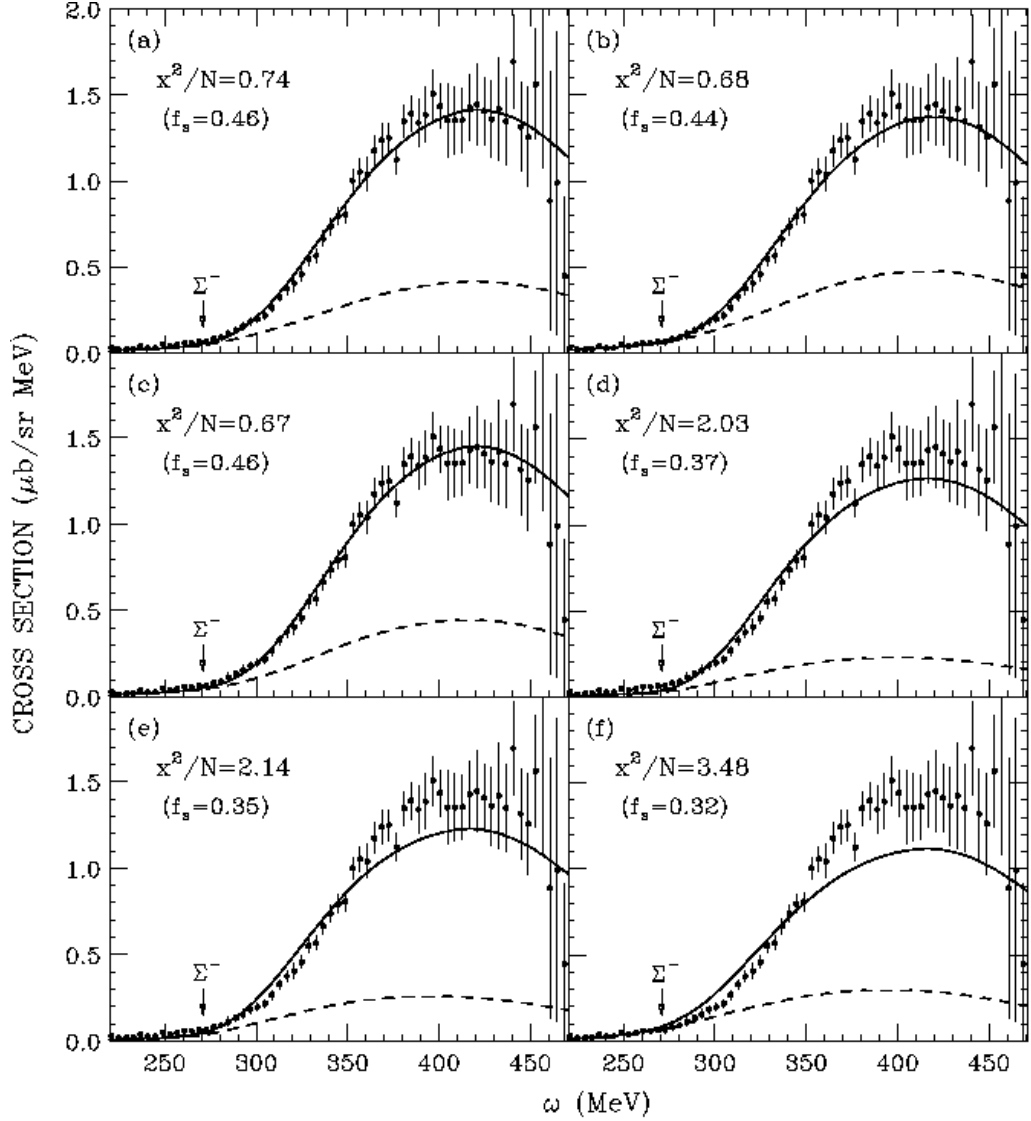}
\caption{Comparison between DWIA calculations~{\protect\cite{HHi05}}
and the measured $^{28}{\rm Si}(\pi^-,K^+)$ spectrum~{\protect\cite{SNA04}} 
using six $\Sigma$-nucleus potentials, (a)-(c) with inner repulsion,
(d)-(f) fully attractive. The solid and dashed curves denote the
inclusive and $\Lambda$ conversion cross sections, respectively.
Each calculated spectrum was normalized by a fraction $f_s$. The arrows
mark the ${\Sigma^-} - {^{27}{\rm Al}_{\rm g.s.}}$ threshold at
$\omega = 270.75$~MeV.}
\label{fig:haradaqf} 
\end{figure} 

A more straightforward information on the nature of the $\Sigma$-nuclear 
interaction has been provided by recent measurements of inclusive 
$(\pi^-,K^+)$ spectra on medium to heavy nuclear targets at KEK~\cite{NSA02,
SNA04}. The inclusive $(\pi^-,K^+)$ spectra on Ni, In and Bi are shown in 
Fig.~\ref{fig:ninbispec} together with a fit using Woods-Saxon potentials 
with depths $V_0 = 90$ MeV for the (repulsive) real part and $W_0=-40$ MeV 
for the imaginary part. These and other spectra measured on lighter targets 
suggest that a strongly {\it repulsive} $\Sigma$-nucleus potential is required 
to reproduce the shape of the inclusive spectrum, while the sensitivity to 
the imaginary (absorptive) component is secondary. The favored strength of 
the repulsive potential in this analysis is about 100 MeV, of the same order 
of magnitude reached by the DD $\Sigma^-$ atomic fit potential shown in 
Fig.~\ref{fig:ddsig} as it `enters' the nucleus inward. The general level of 
agreement in the fit shown in Fig.~\ref{fig:ninbispec} is satisfactory, but 
there seems to be a systematic effect calling for more repulsion, 
the heavier is the target. 
We conclude that a strong evidence has been finally established for the 
repulsive nature of the $\Sigma$-nucleus potential. 

More sophisticated theoretical analyses of these KEK $(\pi^-,K^+)$ spectra 
\cite{KFW04,KFW06,HHi05,HHi06} have also concluded that the $\Sigma$-nuclear 
potential is repulsive within the nuclear volume, although they yield 
a weaker repulsion in the range of 10-40 MeV. An example of a recent 
analysis of the Si spectrum is shown in Fig.~\ref{fig:haradaqf} from 
Ref.~\cite{HHi05} where six different $\Sigma$-nucleus potentials are tested 
for their ability within the Distorted Wave Impulse Approximation (DWIA) 
to reproduce the measured $^{28}{\rm Si}(\pi^-,K^+)$ 
spectrum~\cite{SNA04}. This particular DWIA version was tested 
on the well understood $^{28}{\rm Si}(\pi^+,K^+)$ quasi-free $\Lambda$ 
hypernuclear spectrum also taken at KEK with incoming pions of the same 
momentum $p_{\rm lab} = 1.2$~GeV/c. Potential (a) is the DD, type A' potential 
of Ref.~\cite{BFG94b}, (b) is one of the RMF potentials of Ref.~\cite{MFG95}, 
that with $\alpha_{\omega} = 1$, and (c) is a local-density approximation 
version of a $G$ matrix constructed from the Nijmegen model F. These three 
potentials are repulsive within the nucleus but differ considerably there 
from each other. Potentials (d)-(f) are all attractive within the nucleus, 
with (f) being of a $t_{\rm eff}\rho$ form. All of the six potentials are 
attractive outside the nucleus, as required by fits to the `attractive' 
$\Sigma^-$ atomic level shifts. The figure shows clearly, and judging by the 
associated $\chi^2/{\rm N}$ values, that fully attractive potentials are 
ruled out by the data and that only the {\it `repulsive'} $\Sigma$-nucleus 
potentials reproduce the spectrum very well, but without giving preference to 
any of these potentials (a)-(c) over the other ones in this group. 
It was shown by Harada and Hirabayashi~\cite{HHi06}, furthermore, 
that the $(\pi^-,K^+)$ data on targets with neutron excess, such as 
$^{209}$Bi, also lack the sensitivity to confirm the presence of a sizable 
(repulsive for $\Sigma^-$) isovector component of the $\Sigma$ nucleus 
interaction as found in the $\Sigma^-$-atom fits~\cite{BFG94a,BFG94b,MFG95}.

\begin{acknowledgments}

Special thanks are due to the Directors of Course CLXVII 
`strangeness and spin in fundamental physics' Mauro Anselmino and 
Tullio Bressani, to the Scientific Secretary Alessandro Feliciello 
and to the School Secretary Ms. Barbara Alzani and her crew. This Review 
was supported in part by the Israel Science Foundation grant 757/05.

\end{acknowledgments}

\end{document}